# Magnetic dipolar modes in magnon-polariton condensates

E. O. Kamenetskii

Microwave Magnetic Laboratory,
Department of Electrical and Computer Engineering,
Ben Gurion University of the Negev, Beer Sheva, Israel

April 20, 2021

**Abstract**

For dipole-carrying excitations observed in a high-quality resonator, strong-coupling modes can appear as composite bosons with the spontaneous formation of quantized vortices in the condensed phase of a polariton fluid. In exciton-polaritons, in particular, it leads to sustained trapping of the emitted photon. In this paper, we show that magnon-polaritons can be realized due to magnon condensation caused by magnetic dipole-dipole interaction. In a quasi-2D ferrite disk placed in a microwave cavity, one observes quantum confinement effects of magnetic-dipolar-mode (MDM) oscillations. These modes, characterized by energy eigenstates with rotational superflows and quantized vortices, are exhibited as spinor condensates. Along with the condensation of MDM magnons in the quasi-2D disk of the magnetic insulator, electric dipole condensation is also observed. At the MDM resonances, transfer between angular momenta in the magnetic insulator and in the vacuum cavity, demonstrates generation of vortex flows with fixed handedness. This indicates unique topological properties of polariton wavefronts. One observes curved wavefronts and effects of supperradiance in microwave structures. In an environment of scattering states of microwave waveguide, EM waves can carry the topological phases of MDM resonances.

## I. INTRODUCTION

An interaction between the 'bare' photon and 'bare' medium dipole-carrying excitation becomes strong enough near the resonance between the light mode and the mode of the medium excitation. At the resonance region, the dispersion curves of these modes transform into two split polaritonic branches showing anticrossing behavior. A strong-coupling regime of polaritons leads to many interesting physical properties. The most discussed types of polaritons are phonon polaritons, exciton polaritons, surface-plasmon polaritons, and magnon polaritons.

Polaritons are bosonic quasiparticles. In a case of exciton polaritons, one observes strong coupling of the exciton with the optical-cavity photon. The spectral response displays mode splittings when the quantum wells and the optical cavity are in resonance. A number of quantum-well resonances were shown experimentally. Classically, the effect can be seen as the normal-mode split of coupled oscillators, the excitons and the electromagnetic field of the microcavity. Quantum mechanically, this is the Rabi vacuum-field splitting of the quantum-well excitons. Due to Bose–Einstein condensation of exciton-polaritons in semiconductor optics, one observes spontaneous phase transitions to quantum condensed phases with superfluidity and vortex formation. A quantized vortex is considered as a topological defect with zero density at its core. Quantization of a vortex in a BEC originates from the single-valuedness of the macroscopic wave function, that is, a change in the phase along an arbitrary close path must be an integral multiple of $2\pi$. At the same time, the $\pi$-rotation effects can be observed. There are the spinor exciton-polariton condensates with half-integer vortices. It is worth noting also that excitons, by virtue of being composite particles made of two fermions, obey bosonic statistics as long as their density is low enough such that they do not overlap. In a small quantum dot, the excitons behave as fermions: we cannot put more than one in the same state. In this limit, we perceive the fermionic nature of the constituents [1 – 6].

In microwaves, we are witnesses that long-standing research in coupling between electrodynamics and magnetization dynamics noticeably reappear in recent studies of strong magnon-photon interaction [7 – 11]. In a small ferromagnetic particle, the exchange interaction can lead to the fact that a very large number of spins to lock together into one macrospin with a corresponding increase in oscillator strength. This results in strong enhancement of spin-photon coupling. In a structure of a microwave cavity with a yttrium iron garnet (YIG) sphere inside, the avoided crossing in the microwave reflection spectra verifies strong coupling between the microwave photon and the macrospin magnon. In these studies, the Zeeman energy is defined by a coherent state of the macrospin-photon system when a magnetic dipole is in its antiparallel orientation to the cavity magnetic field. Together with an analysis of the strong coupling of the electromagnetic modes of a cavity with the fundamental Kittel modes, coupling with non-uniform modes – the Walker modes – in a YIG sphere was considered. In the microwave experiments, identification of the Walker modes in the sphere was made based an effect of overlapping between the cavity and spin waves due to relative symmetries of the fields [12, 13]. Nevertheless, the experimentally observed effects of strong magnon-photon interaction, cannot be described properly in terms of a macrospin-photon coupling process. In a view of these aspects, the theory based on solving coupled Maxwell and Landau-Lifshitz-Gilbert equations without making the conventional magnetostatic approximation have been suggested [14, 15]. Currently, the studies of strong magnon-photon interaction are integrated in a new field of research called cavity spintronics (or spin cavitronics) [16].

In magnetic insulators, magnons – quantized fluctuations of the magnetic order parameter – can undergo Bose-Einstein condensation. In this quasi-equilibrium condensation, one observes the magnon-conserving scattering processes, before they dissipate energy to the lattice and relax. In experiments on quasi-equilibrium *magnon condensation* in solid-state magnetic insulators, the excitation of magnons is achieved by microwave pumping [17]. However, as far as we know, *magnon-polariton condensates* – quasiparticles consisting of a superposition of magnon and microwave photons – have not yet been observed. Moreover, as we can see, the possibility of observing such magnon-polariton condensates with superfluidity and vortex formation has not even been discussed in the literature.

The coupling strength in the magnon-photon system is proportional to the probability of conversion of a photon to a magnon and vice versa. An effective way for strong coupling is to confine both magnons and photons to a small (subwavelength) resonant region. Long-range spin transport in magnetic insulators demonstrates that the dipolar interactions alone generate coherent spin waves on the scales that are much larger than the exchange-interaction scales and, at the same time, much smaller than the electromagnetic-wave scales. However, it is generally accepted that magnon condensation on the scales of magnetic dipole-dipole interactions in magnetic insulators is difficult to realize. It is discussed that dipolar interactions destroy spin superfluidity and that the anisotropy resulting from the dipolar interaction makes microwave detection of such condensed magnons very difficult [18, 19].

At the same time, it should be noted that numerous studies show that in the structures of an ultra-cold atomic gas, the role of magnetic dipole-dipole interaction is significant for trapping a Bose-Einstein condensate in a ring geometry with inducing rotational superflow [20, 21]. The long-range nature and anisotropy of the dipolar interaction poses challenging questions concerning the stability and superfluidity of the BEC. The dipole interaction couples the spin and orbital angular momenta so that an initial magnetization of the system causes the gas to rotate mechanically. The dipole interaction conserves not spin or orbital angular momentum separately, but the sum of them. When the trapping potential is axisymmetric, the projected total angular momentum onto the symmetry axis is a good quantum number. Due to magnetic dipole-dipole interaction in a spin-polarized structure, one has the spin vortex formation. An important difference from the Einstein-de Haas [22] and Barnett [23] effects in solid-state magnets is that in atomic-gas BEC the spin angular momentum is transferred not to rigid-body rotation but to spin vortices having the angular momentum. The long-range nature of the magnetic dipole-dipole interaction favors spontaneous formation of topological spin textures. In atomic-gas structure with strong magnetic dipole-dipole interaction, one can observe a chiral-vortex BEC. There is a spinor BEC. In such a state, the spin configurations of the right-handed and left-handed screws are transformed into each other under space inversion. When the rotation frequency of rapidly rotating atomic Bose gases is close to the harmonic trap frequency, the atoms are strongly correlated and one observers fermionization of the system composed of bosons [24, 25]. All these aspects concerning magnetic dipole-dipole interaction in ultra-cold atomic gas structures are important in our studies of magnon-polariton condensates. We also observe the dipolar and spinor BEC with the effects of spontaneous formation of topological spin texture, vortex nucleation and breaking of rotational and chiral symmetries. However, in our case, there are quite different phenomena observed at room temperature in a non-rotating ferrite sample.

The induced anisotropy and nonlocality of the dipole-dipole interaction in the YIG sample make the character of magnetic oscillations very sensitive to geometry of the sample. Therefore, the shape of such a YIG sample should strongly affect the magnon-photon characteristics. Interaction of the Walker-mode oscillations in a YIG sphere with microwaves were observed and analyzed long ago [26, 27]. Such oscillations in a thin ferrite disk had also been studied [28]. What is striking is the fact that although only a few absorption peaks are visible for the ferrite sphere, magnetic resonances in a thin ferrite disk are represented by a very rich spectrum of sharp peaks. A normally magnetized quasi-2D ferrite disk embedded in a microwave cavity appears as an open high-Q resonator with multiresonant spectra of both, Fano and Lortenzian, types of the Walker-mode oscillation peaks [28 – 30]. Initially, such a strong difference in the spectra of a ferrite sphere and a thin ferrite disk was explained by the effect of non-uniformity of internal DC magnetic fields in disk-shaped samples [29, 31]. Later, however, it was shown that the main mechanism underling the Walker modes in a quasi-2D ferrite disk is the unique spectral characteristics of these oscillations and that non-uniformity of internal DC magnetic fields plays an auxiliary role in the analysis [32, 33]. This is a demonstration of fundamentally important properties when a combination of both a specific geometric shape and a material structure is considered. A suitable theory of the interaction of microwave photons with subwavelength ferrite samples should be based on a correct spectral analysis of Walker modes. In a quasi-2D ferrite disk, the quantized forms of the collective matter oscillations – called magnetic-dipolar modes (MDMs) – were found to be quasiparticles with both wave-like and particle-like behaviors, as expected for quantum excitations. The MDM in a quasi-2D ferrite disk is the coalescence of spin-wave magnons into a specific quantum state of matter, which is characterized by a single dipole-dominated wave function [33 – 39].

An analysis of MDM resonances in a quasi-2D ferrite disk is based on postulates about a physical meaning of the magnetostatic-potential (MS-potential) functions $\psi(\vec{r},t)$ as a complex scalar wavefunction, which presumes a long-range phase coherence in magnetic dipole–dipole interactions. Quantum confinement with magnetic dipole-dipole interaction in a sample is exhibited as a spinor Bose–Einstein condensate. The MDM oscillations are characterized by energy eigenstates with rotational superflows and quantized vortices. For every MDM, a scalar MS-potential wavefunctions have a phase singularity in the center of a disk and a non-zero azimuth component of the flow velocity demonstrates the vortex structure. The vortices are guaranteed by chiral edge states on a lateral surface of a quasi-2D ferrite disk at the MDM resonances. The orbital degrees of freedom behaves as if we have fermions in the system, even though our system is composed of bosons. The MDM ferrite-disk resonator is an open quantum system. The near fields in the proximity of the MDM ferrite disk are observed as a composition of rotating electric and magnetic fields. Such MDM-originated fields – called magnetoelectric (ME) fields – carry both spin and orbital angular momentums. ME fields are different from free-space electromagnetic (EM) fields. The pseudoscalar helicity parameters and power-flow vortices of ME fields define unique topological properties of MDM oscillations [33 – 39].

The coupling of a subwavelength MDM resonator with a bosonic-field microwave cavity is far from a trivial problem. In this paper, we show that magnon polaritons can be realized due to magnon condensation caused by dipole-dipole interaction in a quasi-2D ferrite disk. This is considered as a main mechanism of strong coupling of the MDM with the microwave-cavity photon. The interaction between MDM ferrite particles and an external EM field is analyzed based on the notion of an anapole moment [36, 40] and helical-mode MS resonances [41], and considering conservation of an angular momentum in an entire microwave structure [38, 42]. The topological properties of microwave-cavity fields arise due to the MDM states. The eigenstates of the system of a ferrite disk and microwave radiation field are mixtures of photons and MDM magnons. When a MDM ferrite disk is placed in a microwave cavity, one observes mixing space and time components and the effects of spacetime curvature of the cavity fields. For modeling magnon-polariton condensates in a microwave cavity, we propose the concept of double-helix resonances creates by ME photons. It is important to note that for MDM resonances in magnetic insulators, along with magnon condensation, we also have electric dipole condensation. Electric dipoles in a ferrite disk are described by a vector order parameter and therefore exhibits spontaneous symmetry breaking. In microwave waveguide, we observe rotational superradiant scattering of microwave photons by MDM vortices. In an environment of scattering states of microwave waveguide, EM waves can carry the topological phases of MDM resonances.

The paper is organized as follows. Starting with a brief review of our previously published theory on quantum confinement in magnetic structures with dipole-dipole interaction, in Sec. II we show that dipole-dipole condensed structures will be realizable when MDM magnons are strongly trapped by rotational superflow in a ring geometry of a quasi-2D ferrite disk. In Sec. III, we argue that MDM magnons in a quasi-2D ferrite disk are condensed fermions. At the MDM resonances, one observes half-integer quantization of the orbital angular momentum. There exist the right- and left-handed magnetostatic helical waves for a given direction of a bias magnetic field. In Sec. IV, we analyze the electric field structures at the MDM resonances. We show that along with the condensation of MDM magnons in the quasi-2D disk of the magnetic insulator, electric dipole condensation is also observed. MDM resonator is an open quantum system. Near fields of such a MDM resonator (called ME fields) have unique topological properties. In Sec. V, we analyze the symmetry properties of ME near fields. At the MDM resonance, the internal angular momenta (local precessional and collective orbital) of the magnetization fields in a ferrite disk, must be balanced with the external (precessional and orbital) angular momenta of the currents on a metallic waveguide wall. This results in appearance of the Casimir torque. The problems of cavity quantum electrodynamics with MDM disks, discussed in Sec. VI, is determined by the Casimir torque effect. In section VII, we consider rotational superradiant scattering of microwave photons by MDM vortices. This is due to amplification of an amplitude of the incident electromagnetic wave by extracting energy from the collective motion of orbitally rotating spinning magnetic dipoles. In a structure of microwave waveguide with an embedded MDM ferrite disk, one observes curved wavefronts. Finally, in Sec. VIII, we present discussion and conclusion.

## II. CONDENSATION OF MAGNONS IN CONFINED STRUCTURES WITH MAGNETIC DIPOLE-DIPOLE INTERACTION

The dipole-dipole interaction provides us with a long-range mechanism of interaction, where a magnetic medium is viewed as a continuum. No small-scale microscopic effects are taken into account, and in the analysis of local fluctuations of the magnetization caused by magnetic dipole waves, small-scale exchange-interaction spin waves are beyond the consideration. The dipole-dipole collective excitation in a small ferrite sample can be determined by a MS-potential wave function $\psi(\vec{r},t)$. It is assumed that a wave function $\psi(\vec{r},t)$ represents the state of the system. It completely describes the behavior of condensed magnons – microscopically precessing localized spins – in a ferrite disk at the MDM resonances.

The MDM-resonance spectral solutions for function $\psi(\vec{r},t)$ obtained from the second-order differential equation – the Walker equation – are constructed in accordance with basic symmetry considerations for the sample geometry. For an open quasi-2D ferrite disk normally magnetized along the $z$ axis, we can use separation of variables [34]. In a cylindrical coordinate system $(z, r, \theta)$, the solutions are represented as

$$\psi_{p,\nu,q} = A_{p,\nu,q} \xi_{p,\nu,q}(z) \tilde{\eta}_{\nu,q}(r,\theta), \qquad (1)$$

where $A_{p,\nu,q}$ is a dimensional amplitude coefficient, $\xi_{p,\nu,q}(z)$ is a dimensionless function of the MS-potential distribution along $z$ axis, and $\tilde{\eta}_{\nu,q}(r,\theta)$ is dimensionless membrane function. The membrane function $\tilde{\eta}$ is defined by a Bessel-function order $\nu$ and a number of zeros of the Bessel function corresponding to a radial variations $q$. The dimensionless "thickness-mode" function $\xi(z)$ is determined by the axial-variation number $p$. Confinement of MDM oscillations leads to energy quantization. In a quasi-2D ferrite disk, one can formulate the energy eigenstate boundary problem based on the Schrödinger-like equation for scalar-wave eigenfunctions $\psi(\vec{r},t)$ with using the Dirichlet-Neumann boundary conditions. For MDM $n$ ( $n \equiv p,\nu,q$ ), the energy eigenvalue problem is defined by the differential equation [34]

$$\hat{G}_\perp \tilde{\eta}_n = E_n \tilde{\eta}_n, \qquad (2)$$

where $\hat{G}_\perp$ is a two-dimensional (on the disk plane) differential operator. The quantity $E_n$ is interpreted as density of accumulated magnetic energy of mode $n$. This is the average (on the RF period) energy accumulated in a flat ferrite-disk region of unit in-plane cross-section and unit length along $z$ axis. The operator $\hat{G}_\perp$ and quantity $E_n$ are defined as

$$\frac{g_n \mu_0}{4} \mu_n \nabla_\perp^2 \tilde{\eta}_n = E_n \tilde{\eta}_n, \qquad (3)$$

where

$$E_n = \frac{g_n \mu_0}{4} (\beta_n)^2 . \qquad (4)$$

Here $\nabla_\perp^2$ is the two-dimensional (on the circular cross section of a ferrite-disk region) Laplace operator, $g_n$ is a dimensional normalization coefficient (with the unit of dimension $\psi^2$) for mode $n$, and $\beta_n$ is the propagation constant of mode $n$ along the disk axis $z$. The parameter $\mu_n$ (which is a diagonal component of the permeability tensor [43]) should be considered as an eigenvalue. Outside a ferrite $\mu_n = 1$. The operator $\hat{G}_\perp$ is a self-adjoint operator only for negative quantities $\mu_n$ in a ferrite. For self-adjointness of operator $\hat{G}_\perp$, the membrane function $\tilde{\eta}_n(r,\theta)$ must be continuous and differentiable with respect to the normal to lateral surface of a ferrite disk. The homogeneous boundary conditions – the Neumann-Dirichlet (ND) boundary conditions – for the membrane function should be: $(\tilde{\eta}_n)_{r=\mathcal{R}^-} - (\tilde{\eta}_n)_{r=\mathcal{R}^+} = 0$ and $\mu \left( \frac{\partial \tilde{\eta}_n}{\partial r} \right)_{r=\mathcal{R}^-} - \left( \frac{\partial \tilde{\eta}_n}{\partial r} \right)_{r=\mathcal{R}^+} = 0$, where $\mathcal{R}$ is a disk radius. MDM oscillations in a ferrite disk are described by real eigenfunctions: $\tilde{\eta}_{-n} = \tilde{\eta}_n^*$. For modes $n$ and $n'$, the orthogonality conditions are expressed as

$$\int_{S_c} \tilde{\eta}_n \tilde{\eta}_{n'}^* dS = \delta_{nn'} , \qquad (5)$$

where $S_c$ is a square of a circular cross section of a ferrite-disk region and $\delta_{nn'}$ is the Kronecker delta. The spectral problem gives the energy orthogonality relation for MDMs: $(E_n - E_{n'}) \int_{S_c} \tilde{\eta}_n \tilde{\eta}_{n'}^* dS = 0$. Since the space of square integrable functions is a Hilbert space with a well-defined scalar product, we can introduce a basis set. A dimensional amplitude coefficient we write as $A_n = c' a_n$, where $c'$ is a dimensional unit coefficient and $a_n$ is a normalized dimensionless amplitude. The normalized scalar-wave membrane function $\tilde{\eta}$ can be represented as $\tilde{\eta} = \sum_n a_n \tilde{\eta}_n$. The amplitude is defined as $|a_n|^2 = \left| \int_{S_c} \tilde{\eta} \tilde{\eta}_n^* dS \right|^2$. The mode amplitude can be interpreted as the probability to find a system in a certain state $n$. Normalization of membrane function is expressed as $\sum_n |a_n|^2 = 1$ [33 – 39].

As it was noted above, observation of condensed states of dipolar magnetization is hardly possible, since spin superfluidity is destroyed on the scales of magnetic dipole-dipole interactions. It means that dipole-carrying excitations in a high-quality confined ferrite-disk structure cannot be identified as the quantum states of microscopic short-range excitations stabilized by the condensate phase. Nonetheless, following the studies with ultra-cold atomic gas structures [20, 21], one can assume that microwave detection of dipole-dipole condensed structures will be realizable when MDM magnons are strongly trapped by *rotational superflow* in a ring geometry of a quasi-2D ferrite disk. In a pattern of the rotating magnetization, the dipole interaction couples the spin and orbital angular momenta. The spin angular momentum is converted into the orbital angular momentum of spin-vortex state via the spin-orbit interaction. The sum of the spin or orbital angular momenta is conserved. For energetically favorable angular velocity, the vortex is formed. The magnetic-dipolar condensate is shown to exhibit a novel ground state, which has a net orbital angular momentum with broken chiral symmetry. The effect of orbitally rotating MDM magnons in a quasi-2D ferrite disk was observed both numerically and experimentally. Theoretically, it can be explained based on an analysis of boundary conditions.

In solving boundary value problems for MS resonances, one encounters some non-trivial questions when using boundary conditions. As is known, in solving a boundary value problem that involves the eigenfunctions of a differential operator, the boundary conditions must be in a definite correlation with the type of this differential operator [44, 45]. In an analysis of MDM resonances in a ferrite disk, we used the homogeneous ND boundary conditions, which mean continuity of the MS wave functions together with continuity of their first derivatives on the sample boundaries. Only in this case we have a complete set of orthogonal basis functions and thus the field expansion in terms of orthogonal MS-potential functions can be employed. Surprisingly, this fact, well known from text books of functional analysis, is not taken into consideration in numerous publications regarding MS resonances in ferrite samples. In these works, the solutions were obtained based on the Walker's second-order differential equation for MS wave functions and the boundary conditions for *dynamic magnetization*. Certainly, these boundary conditions cannot be considered as suitable boundary conditions for the Walker-equation MS magnons. Different types of orthogonality relations derived for dynamic magnetization of long-wavelength (or magnetostatic) magnons [46 – 48] are not appropriate for self-adjointness of the Walker-equation differential operator. This question also arises in some general formulations of the boundary value problem for differential operators of long-range magnons in the continuum limit. [49]. In addition, it should be noted that the boundary conditions for dynamic magnetization are not the EM boundary conditions [50].

However, the ND boundary condition, used in the above analysis of MDM oscillations, are not the EM boundary conditions as well. While the considered above ND boundary conditions are the so-called essential boundary conditions, the EM boundary conditions are the natural boundary conditions [44]. For the EM boundary conditions on lateral surface of a ferrite disk, it is necessary to have both the continuity of a membrane function $\tilde{\eta}(r,\theta)$ and a radial component of the magnetic flux density $B_r = \mu_0 \left( \mu \frac{\partial \tilde{\eta}}{\partial r} + \mu_a \frac{\partial \tilde{\eta}}{\partial \theta} \right)$. Here $\mu$ and $\mu_a$ are diagonal component and off-diagonal components of the permeability tensor [43]. With the EM boundary conditions, it becomes evident that the membrane function $\tilde{\eta}$ must not only be continuous and differentiable with respect to a normal to the lateral surface, but, because of the presence of a gyrotropy term, be also differentiable with respect to a tangent to this surface. This means the presence of an azimuth magnetic field on the border circle. In this case, membrane functions $\tilde{\eta}$ cannot be considered as single-valued functions, and the question arises of the validity of the energy orthogonality relation for MS-wave modes.

To restore the ND boundary conditions and thus the completeness of eigenfunctions $\tilde{\eta}$, we need introducing a certain surface magnetic current $j_s^{(m)}$ circulating on a lateral surface of the disk. This current must compensate the term $\left( i\mu_a \frac{1}{r} \frac{\partial \tilde{\eta}}{\partial \theta} \right)_{r=\mathcal{R}^-}$ in the expression of the radial component of the magnetic flux density [36, 37, 40]. Evidently, for a given direction of a bias magnetic field (that is, for a given sign of $\mu_a$), there are two, clockwise and counterclockwise, quantities of a circulating magnetic current. The current $j_s^{(m)}$ is defined by the velocity of an irrotational border flow. This flow is observable via the circulation integral of the gradient $\vec{\nabla}_\theta \delta = \frac{1}{\mathcal{R}}\left(\frac{\partial \delta_\pm}{\partial \theta}\right)_{r=\mathcal{R}} \vec{e}_\theta$, where $\delta_\pm$ is a double-valued edge wave function on a contour $\mathcal{L} = 2\pi\mathcal{R}$. On a lateral surface of a quasi-2D ferrite disk, one can distinguish two different functions $\delta_\pm$, which are the counterclockwise and clockwise rotating-wave edge functions with respect to a membrane function $\tilde{\eta}$. The entire cycle of $2\pi$ rotation of wavefunction $\delta_\pm$ corresponds to the $\pi$-shift of membrane function $\tilde{\eta}$. As a result, one has the eigenstate spectrum of MDM oscillations with topological phases accumulated by the edge wave function $\delta$. A circulation of gradient $\vec{\nabla}_\theta \delta$ along the contour $\mathcal{L} = 2\pi\mathcal{R}$ gives a non-zero quantity when an azimuth number is a quantity divisible by $\frac{1}{2}$. A line integral around a singular contour $\mathcal{L}$:

$\frac{1}{\mathfrak{R}}\oint_\mathcal{L}\left(i\frac{\partial \delta_\pm}{\partial \theta}\right)(\delta_\pm)^* d\mathcal{L} = \int_0^{2\pi}\left[\left(i\frac{\partial \delta_\pm}{\partial \theta}\right)(\delta_\pm)^*\right]_{r=\mathfrak{R}} d\theta$ is an observable quantity. Each MDM is quantized to a quantum of an emergent electric flux. There are the positive and negative eigenfluxes. These different-sign fluxes should be nonequivalent to avoid the cancellation. The Berry mechanism provides a microscopic basis for the surface magnetic current at the interface between gyrotropic and nongyrotropic media. Following the spectrum analysis of MDMs in a quasi-2D ferrite disk one obtains pseudo-scalar axion-like fields and edge chiral magnetic currents. The anapole moment for every mode $n$ is calculated as [36, 37, 40]:

$$a_\pm^{(e)} \propto \mathcal{R}\int_0^d \oint_\mathcal{L}\left[\vec{j}_s^{(m)}(z)\right]_\theta \cdot d\vec{l}dz. \tag{7}$$

One observes the anapole moment due to nonzero circulation of magnetization on a lateral surface of a quasi-2D ferrite disk at the MDM resonances. The evidence for such a circulation will be discussed below.

For the EM boundary conditions, the solution for MS-potential wave function is written now as

$$\psi_{p,\nu,q} = C_{p,\nu,q}\xi_{p,\nu,q}(z)\tilde{\varphi}_{\nu,q}(r,\theta), \tag{8}$$

where $C_{p,\nu,q}$ is a dimensional amplitude coefficient and $\tilde{\varphi}_{\nu,q}$ is a membrane function. The MS-potential membrane wave function is expressed as: $\left(\tilde{\varphi}_\pm\right)_{r=\mathfrak{R}^-} = \delta_\pm\left(\tilde{\eta}\right)_{r=\mathfrak{R}^-}$. For a given direction of a bias magnetic field, the membrane function of any mode $n$ ($n \equiv p,\nu,q$), is presented as a two-component sprinor:

$$\tilde{\varphi}_n(\vec{r},\theta) = \tilde{\eta}_n(\vec{r},\theta) \begin{bmatrix} e^{-\frac{1}{2}i\nu_n\theta} \\ e^{+\frac{1}{2}i\nu_n\theta} \end{bmatrix}. \tag{9}$$

MDM oscillations with ND boundary conditions, described by the functions of Eq. (1), we conditionally call as *G*-modes. MDM oscillations with EM boundary conditions, described by the functions of Eq. (8), are called as *L*-modes [37, 41]. Since $|\tilde{\varphi}_n| = |\tilde{\eta}_n|$, we have the same energy eigenstates for the *G*- and *L*-modes. A comparative analysis of these two types of modes shows that the half-integer quantization of the orbital angular momentum in a quasi-2D ferrite disk is observed due to the topological phases accumulated by the edge wave functions. The existence of topological-phase effects with the half-integer quantization of the orbital angular momentum is a well-known phenomenon in structures with reduced dimensions. As other examples, we can refer to works on twisted light with half-integer angular momentum [51] and half-integer orbital angular momenta in circular quantum dots [52].

## III. MDM MAGNONS AS CONDENSED FERMIONS

A membrane function $\tilde{\varphi}_n$, represented as a two-component spinor, can be viewed a membrane eigenfunction $\tilde{\eta}_n$ bouncing between two disk planes and rotates around the disk axis. With separation of variables, one can consider the components of the power flow density (current density). For mode $n$, there are the axial $(\vec{\mathcal{J}}_\parallel)_n$ and in-plane $(\vec{\mathcal{J}}_\perp)_n$ components. For the total current density, we have [36, 39]:

$$\vec{\mathcal{J}}_n = (\vec{\mathcal{J}}_\parallel)_n + (\vec{\mathcal{J}}_\perp)_n, \tag{10}$$

where $(\vec{\mathcal{J}}_\parallel(z))_n = \frac{i\omega}{4}\left[\psi_n(\vec{B}_\parallel^*)_n - \psi_n^*(\vec{B}_\parallel)_n\right]$ and $(\vec{\mathcal{J}}_\perp(\vec{r},\theta))_n = \frac{i\omega}{4}\left[\psi_n(\vec{B}_\perp^*)_n - \psi_n^*(\vec{B}_\perp)_n\right]$. The axial component $(\vec{\mathcal{J}}_\parallel)_n$ is directed along $z$ axis:

$$(\vec{\mathcal{J}}_z)_n = -\frac{i\omega}{4}\mu_0|C_n|^2|\tilde{\eta}|^2\left[\xi_n\left(\frac{\partial \xi_n}{\partial z}\right)^* - \xi_n^*\frac{\partial \xi_n}{\partial z}\right]\vec{e}_z, \tag{11}$$

For the in-plane component $(\vec{\mathcal{J}}_\perp)_n$, we have only azimuthal constituent:

$$(\vec{\mathcal{J}}_\theta)_n = -\frac{i\omega}{4}|C_n|^2|\xi_n|^2\mu_0\left\{\tilde{\varphi}_n\left(-i\mu_a\frac{\partial \tilde{\varphi}_n}{\partial r} + \mu\frac{1}{r}\frac{\partial \tilde{\varphi}_n}{\partial \theta}\right)^* - \tilde{\varphi}_n^*\left(-i\mu_a\frac{\partial \tilde{\varphi}_n}{\partial r} + \mu\frac{1}{r}\frac{\partial \tilde{\varphi}_n}{\partial \theta}\right)\right\}\vec{e}_\theta. \tag{12}$$

Non-zero quantities of the power flow circulation (clockwise or counterclockwise) around a circle $L = 2\pi r$, where $0 < r \leq \mathcal{R}$, are vortices with cores at the disk center. In a disk of unit thickness and for a ring of unit width, we have an angular momentum generated by a circulating flow of energy in the wave field:

$$\vec{L}_z = \oint_L \vec{r} \times \frac{1}{r} \left(\vec{\mathcal{J}}_\theta\right)_n dl .$$ (13)

The presence of quantized power-flow vortices of the modes means that some quantized angular gradients are required to switch the process on. This is an evidence of the persistent power flow (persistent current) in the disk circumference. In an attempt to obtain a concrete physical picture of the spin of the electron, Ohanian [53] argues that the angular momentum is carried by a circulating flow of energy, suitably constructed for the spin field. The spin may be related as an angular momentum generated by a circulating flow of energy in the wave field. Likewise, the magnetic moment may be regarded as generated by a circulating flow of charge. In our case, we can consider an orbital angular momentum of the MDM ferrite disk as a half-integer internal angular momentum related to a circulating flow of energy. It is worth noting also that we are talking about pseudoangular momentum. While the conservation of angular momentum follows from invariance under rotations of all the constituents of a solid, the conservation of pseudoangular momentum follows from invariance under rotations of fields while keeping the solid fixed [49].

It is important to point out, however, that using the separation of variables in a cylindrical coordinate system, we are far from observing the real physical picture. Due to the strong spin–orbit interaction in the MDM magnetization dynamics, the fields of these oscillations in ferrite disks should have helical structures. The right-left asymmetry of MDM rotating fields is related to helical resonances, in which the phase variations for resonant $\psi$ functions occur in both axial and azimuthal directions. Thus, the correct solutions of a spectral problem for MDMs should be obtained in a helical-coordinate system. The helices are topologically nontrivial structures, and the phase relationships for waves propagating in such structures could be very special. Unlike the Cartesian- or cylindrical-coordinate systems, the helical-coordinate system is not orthogonal, and separating the right-handed and left-handed solutions is admitted. In a helical coordinate system, the solution for MS-potential wave function for MDM oscillations in an open ferrite-disk resonator have four types of waves. These are distinguished as forward right-hand-helix wave, backward right-hand-helix wave, forward left-hand-helix wave, and backward left-hand-helix wave [41]. For every of these types of MS helical waves we can define the power flow density (current density). Since the electron spin precesses counterclockwise around the direction of the bias magnetic field, we can a priori assume that the orbital rotation of the MDM should be observed mainly as counterclockwise relative to the direction of the bias magnetic field. It means that for a given direction of a bias magnetic field, there are only two types of helical waves.

An example of the right- and left-handed MS helical waves for a given direction of a bias magnetic field is shown in Fig. 1. The helices are shown in such a way that the height of one complete helix turn is between two virtual planes. The distance between these planes is equal to the pitch of the helix. In a real structure of an open ferrite disk, the disk thickness is much less than the pitch. It means that for each of these helices inside a ferrite, there is a negligibly small phase variation along $z$ axis.

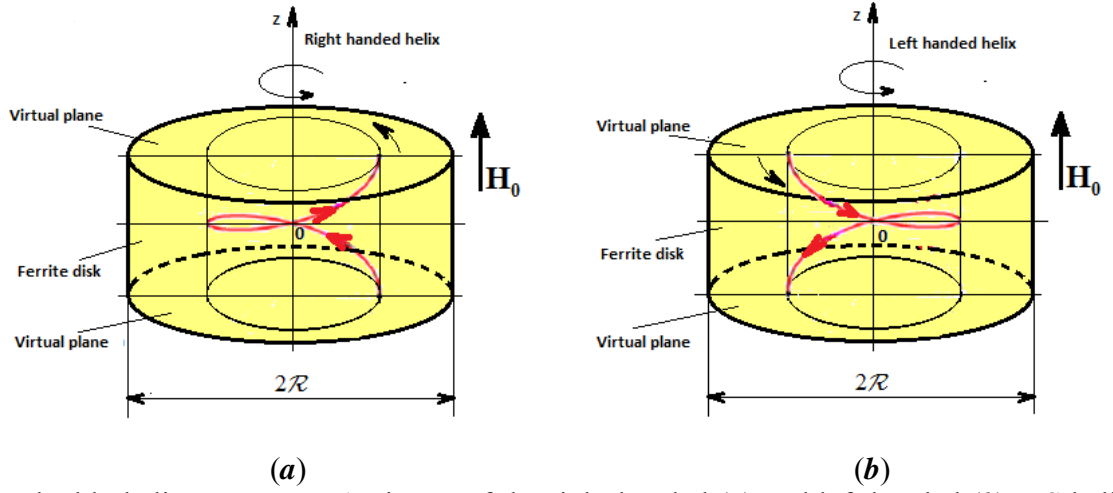

*(a)*  *(b)*

Fig.1. The double-helix resonance. A picture of the right-handed (*a*) and left-handed (*b*) MS helical waves are shown for a given direction of a bias magnetic field. The helices are displayed in such a way that the height of one complete helix turn is between two virtual planes. In a real structure of an open ferrite disk, the disk thickness is much less than the pitch.

Having correctly selected the orbital rotations of the MDM resonances, we distinguish the right-handed (*RH*) and left-handed (*LH*) screws of MS-potential wave functions. However, using our classification of spiral modes, we will take into account that in real samples the disk thickness is much less than the helix pitch. This makes it possible to return to a cylindrical coordinate system taking into account the four topological mode configurations inside the ferrite disk. For a bias magnetic field directed along the *z*-axis or against it, the following modes are distinguished:

$$^\uparrow\psi_n^{(RH)}(\vec{r},\theta,z) = {}^\uparrow C_n^{(RH)}\, {}^\uparrow\tilde{\eta}_n^{(RH)}(\vec{r},\theta)\, e^{-i\left(\frac{1}{2}\nu_n\theta+\beta_n z\right)}, \tag{14}$$

$$^\uparrow\psi_n^{(LH)}(\vec{r},\theta,z) = {}^\uparrow C_n^{(LH)}\, {}^\uparrow\tilde{\eta}_n^{(LH)}(\vec{r},\theta)\, e^{-i\left(\frac{1}{2}\nu_n\theta-\beta_n z\right)}, \tag{15}$$

$$^\downarrow\psi_n^{(LH)}(\vec{r},\theta,z) = {}^\downarrow C_n^{(LH)}\, {}^\downarrow\tilde{\eta}_n^{(LH)}(\vec{r},\theta)\, e^{i\left(\frac{1}{2}\nu_n\theta-\beta_n z\right)}, \tag{16}$$

$$^\downarrow\psi_n^{(RH)}(\vec{r},\theta,z) = {}^\downarrow C_n^{(RH)}\, {}^\downarrow\tilde{\eta}_n^{(RH)}(\vec{r},\theta)\, e^{i\left(\frac{1}{2}\nu_n\theta+\beta_n z\right)}. \tag{17}$$

Here arrows $^{\uparrow\downarrow}$ mean the directions of a bias magnetic field with respect to the *z* axis.

The model described by Eqs. (14)–(17) allows illustrating the general picture of power circulations inside a ferrite disk based on Eqs. (11) and (12). Fig. 2 shows these four screw modes in the cylindrical coordinates. In this picture, the orbital angular momentum is indicated as a red arrow. It is an axial vector directed along a bias magnetic field. The polar vector of the power-flow propagation along the *z* axis is shown by a black arrow. In the case (*a*), the power flow propagates up with the $^\uparrow\psi_n^{(RH)}(\vec{r},\theta,z)$ helical mode and propagates down with the $^\uparrow\psi_n^{(LH)}(\vec{r},\theta,z)$ helical mode. In the case (*b*), the power flow propagates up with the $^{\downarrow}\psi_n^{(LH)}(\vec{r},\theta,z)$ helical mode and propagates down with the $^{\downarrow}\psi_n^{(RH)}(\vec{r},\theta,z)$ helical mode.

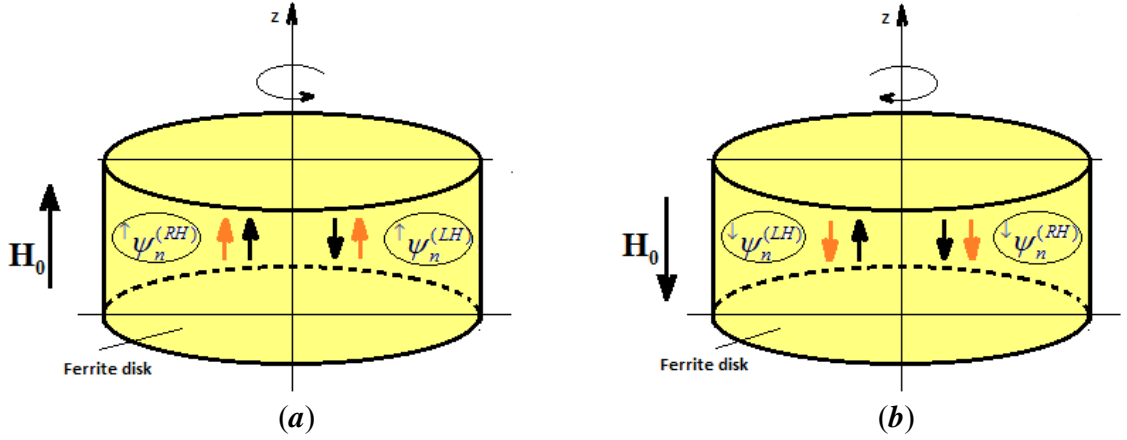

(*a*)                 (*b*)

Fig. 2. Four mode configurations inside the ferrite disk in cylindrical coordinate system at a bias magnetic field directed (*a*) along *z* axis and (*b*) opposite to *z* axis. Orbital angular momentum (shown by a red arrow) is directed along a bias magnetic field. Direction of the wave propagation with respect to is *z* axis is shown by a black arrow.

We can see that while modes $^{\uparrow}\psi_n^{(RH)}(\vec{r},\theta,z)$ and $^{\downarrow}\psi_n^{(RH)}(\vec{r},\theta,z)$ have the "antiparticle" configurations, modes $^{\uparrow}\psi_n^{(LH)}(\vec{r},\theta,z)$ and $^{\downarrow}\psi_n^{(LH)}(\vec{r},\theta,z)$ have the "particle" configurations. The wave functions of the "particle" and "antiparticle" configurations are related by complex conjugation. Since the complex conjugation is related to the time reversal, along with the solution $\psi(\vec{r},t)$, a proper solution is $\psi^*(\vec{r},-t)$. This means that

$$^{\uparrow}C_n^{(RH)} = \left(^{\downarrow}C_n^{(RH)}\right)^*, \quad ^{\uparrow}C_n^{(LH)} = \left(^{\downarrow}C_n^{(LH)}\right)^*, \quad ^{\uparrow}\tilde{\eta}_n^{(RH)} = \left(^{\downarrow}\tilde{\eta}_n^{(RH)}\right)^*, \quad ^{\uparrow}\tilde{\eta}_n^{(LH)} = \left(^{\downarrow}\tilde{\eta}_n^{(LH)}\right)^*. \qquad (18)$$

For the scalar products, we have the following relationships:

$$\left(^{\uparrow}\psi^{(RH)}\right)\left(^{\downarrow}\psi^{(RH)}\right) = \left(^{\uparrow}\psi_n^{(RH)}\right)\left(^{\uparrow}\psi_n^{(RH)}\right)^* \qquad (19)$$

and

$$\left(^{\uparrow}\psi^{(LH)}\right)\left(^{\downarrow}\psi^{(LH)}\right) = \left(^{\uparrow}\psi_n^{(LH)}\right)\left(^{\uparrow}\psi_n^{(LH)}\right)^*. \qquad (20)$$

We assume that for any mode, there are $\left|^{\uparrow\downarrow}C_n^{(RH,LH)}\right|^2 \equiv \left|\frac{1}{2}C_n\right|^2$ and $\left|^{\uparrow\downarrow}\tilde{\eta}_n^{(RH,LH)}\right|^2 \equiv \left|\tilde{\eta}_n\right|^2$. We also assume that $^{\uparrow}\tilde{\eta}_n^{(RH)} = ^{\uparrow}\tilde{\eta}_n^{(LH)}$ and $^{\downarrow}\tilde{\eta}_n^{(RH)} = ^{\downarrow}\tilde{\eta}_n^{(LH)}$. In this case, one can normalize the helical modes according

to the orthogonality conditions (5). It is worth noting that while $^\uparrow\tilde{\varphi}_n^{(RH)}(\vec{r},\theta) = {}^\uparrow\tilde{\varphi}_n^{(LH)}(\vec{r},\theta) = {}^\uparrow\tilde{\eta}_n(\vec{r},\theta)e^{-i\frac{1}{2}\nu_n\theta}$ and $^\downarrow\tilde{\varphi}^{(RH)}(\vec{r},\theta) = {}^\downarrow\tilde{\varphi}^{(LH)}(\vec{r},\theta) = {}^\downarrow\tilde{\eta}_n(\vec{r},\theta)e^{+i\frac{1}{2}\nu_n\theta}$, there are $^\uparrow\psi_n^{(RH)}(\vec{r},\theta,z) \neq {}^\uparrow\psi_n^{(LH)}(\vec{r},\theta,z)$ and $^\downarrow\psi_n^{(RH)}(\vec{r},\theta,z) \neq {}^\downarrow\psi_n^{(LH)}(\vec{r},\theta,z)$. At a given direction of a bias magnetic field, for helical waves of equal amplitude traveling in opposing directions along $z$ axis, there is on average no net propagation of energy. At the same time, circulations of orbital fluxes of energy of the helical modes produce orbital angular momenta directed along a bias magnetic field. Considering a disk with unit thickness, we have for a bias magnetic field directed along $z$ axis:

$$^\uparrow\vec{L}_z^{(RH,LH)} = -\frac{i\omega}{4}|C_n|^2|\xi_n|^2\mu_0\int_0^{\mathcal{R}}\oint_L \vec{r}\times\left[{}^\uparrow\tilde{\varphi}_n^{(RH,LH)}\left(-i\mu_a\frac{\partial\, ^\uparrow\tilde{\varphi}_n^{(RH,LH)}}{\partial r}+\mu\frac{1}{r}\frac{\partial\, ^\uparrow\tilde{\varphi}_n^{(RH,LH)}}{\partial\theta}\right)^*-\right.$$
$$\left.\left({}^\uparrow\tilde{\varphi}_n^{(RH,LH)}\right)^*\left(-i\mu_a\frac{\partial\, ^\uparrow\tilde{\varphi}_n^{(RH,LH)}}{\partial r}+\mu\frac{1}{r}\frac{\partial\, ^\uparrow\tilde{\varphi}_n^{(RH,LH)}}{\partial\theta}\right)\right]\cdot\vec{e}_l\,dldr \quad (21)$$

For a bias magnetic field directed opposite to $z$ axis, we have a similar relationships for $^\downarrow\vec{L}_z^{(RH,LH)}$.

The *RH* and *LH* helical modes are topologically different structures. They do not eliminate each other. On the cross-section plane of the disk, there is a two-mode coflowing topological circulation turning over a regular-coordinate angle $2\pi$. Such a circulation can be clearly viewed in Fig. 1 as interweaving of two types of rotating helical MS modes on the $z = 0$ plane. Considering the case of a quasi-2D ferrite disk, we also have two-mode coflowing topological circulations turning over a regular-coordinate angle $2\pi$. Such topological circulations are for $^\uparrow\psi_n^{(RH)}(\vec{r},\theta,z)$ and $^\uparrow\psi_n^{(LH)}(\vec{r},\theta,z)$ helical modes. The same situation takes place for $^\downarrow\psi_n^{(LH)}(\vec{r},\theta,z)$ and $^\downarrow\psi_n^{(RH)}(\vec{r},\theta,z)$ helical modes. The pictures of a double-helix-resonance circulations on the cross-section plane $z = 0$ of a thin ferrite disk are shown on Fig. 3 for two directions of the bias magnetic field.

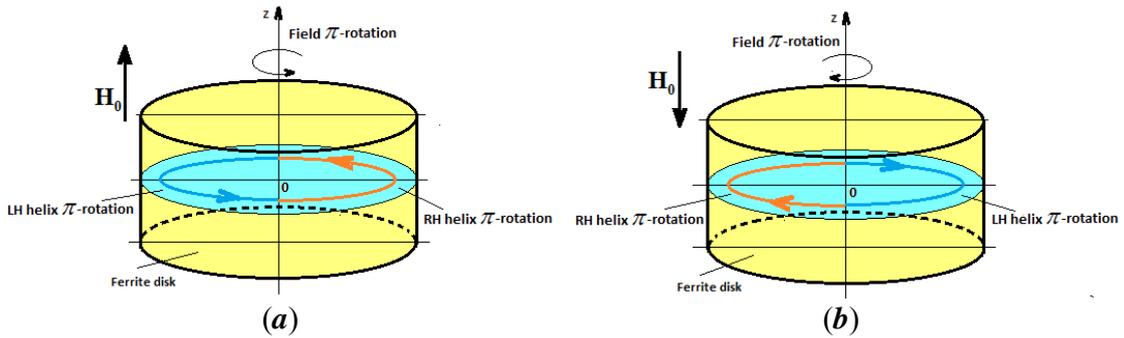

(*a*)          (*b*)

Fig. 3. Illustration of the result of interweaving of two types of rotating helical MS modes on the $z = 0$ plane for a quasi-2D ferrite disk. Assuming the dynamic-phase $\pi$-rotation of the RF EM field, we observe a coflowing turning over a regular-coordinate angle $2\pi$ due to $\pi$-rotation of the *RH* and *LH* helical magnetostatic modes.

The function $\psi(\vec{r},t)$ allows analyzing fields in our continuum model. We can consider this scalar wave function as a generating function in the sense that its partial derivatives generate equations that determine field dynamics of MDM resonances in a ferrite disk. When the spectral problem for the MS-potential scalar wave function $\psi(\vec{r},t)$ is solved, the magnetic field inside a ferrite sample is defined as $\vec{H} = -\vec{\nabla}\psi$. Distribution of magnetization in a ferrite disk is found as $\vec{m} = -\ddot{\chi} \cdot \vec{\nabla}\psi$, where $\ddot{\chi}$ is the susceptibility tensor of a ferrite [43]. Knowing the RF magnetization inside the disk, we can find the outside magnetic field of a MDM by integration of the equation

$$\vec{\nabla} \cdot \vec{H} = -\vec{\nabla} \cdot \vec{m} \tag{22}$$

over a volume and a surface of a ferrite disk. In this case, we have for the magnetic field outside a ferrite disk:

$$\vec{H}_{outside}(\vec{r}) = \frac{1}{4\pi}\left( \int_V \frac{(\vec{\nabla}' \cdot \vec{m}(r'))(\vec{r}-\vec{r}')}{|\vec{r}-\vec{r}'|^3} dV' - \int_S \frac{(\vec{n}' \cdot \vec{m}(r'))(\vec{r}-\vec{r}')}{|\vec{r}-\vec{r}'|^3} dS' \right), \tag{23}$$

where $V$ and $S$ are a volume and a surface of a ferrite sample, respectively. Vector $\vec{n}'$ is the outwardly directed normal to surface $S$. In Eq. (23), we assume that the magnetization fall abruptly to zero at the surface $S$. Application of the divergence theorem to $\vec{\nabla} \cdot \vec{m}$ shows that in such an approximation there is an effective magnetic surface charge density:

$$\sigma^{(m)} = \vec{n}' \cdot \vec{m}. \tag{24}$$

Because of the time-reversal symmetry breaking, resulting from the precessional motion of the magnetization vector about the $z$ axis, the surface magnetic charges cannot cause two, clockwise (CW) *and* counter clockwise (CCW), azimuthal magnetic currents. For the given direction of bias magnetic field, we may have only CW *or* CCW induced magnetic current, which passes over a regular-coordinate angle $\pi$ at the time phase of $\pi$. However, the two screw-mode field configurations at every given direction of a bias magnetic field, provide us with the possibility to have rotational symmetry by a turn over a regular-coordinate angle $2\pi$ at the $\pi$-shift of a dynamic phase of the external RF EM field. In this case, coflowing topological magnetic currents turning over a regular-coordinate angle $2\pi$ appear. The gradient of twisting angle plays the role of the phase gradient.

The magnetization distribution in a quasi-2D ferrite disk at the MDM resonances were analyzed analytically and numerically in Refs. [42, 54, 55]. The precessing magnetization dipoles of orbitally rotating MDMs fit into the circular boundary of the ferrite disk with azimuth anisotropy. To understand the physics of this mechanism, let us consider the situation when the bias magnetic field is directed along the $z$ axis. On this occasion, topological mode configurations inside the ferrite disk in cylindrical coordinate system are described be Eqs. (14) and (15). We assume that the azimuth number $\nu_n = 1$. In this case, the membrane function $\tilde{\eta}_n(r,\theta)$ has a dipolar configuration with respect to an azimuth coordinate [33]. In Fig. 4, we conventionally represent this dipolar configuration as two semicircles of different colors. The magnetization dipoles inside and on the lateral border of a disk for the main mode of the MDM spectrum are shown schematically with blue arrows in Fig. 4. The lengths and widths of these arrows are related to the magnitude of magnetization. The phases $\omega t = 0, \frac{\pi}{2}, \pi$ are the dynamic phases of the external RF EM field.

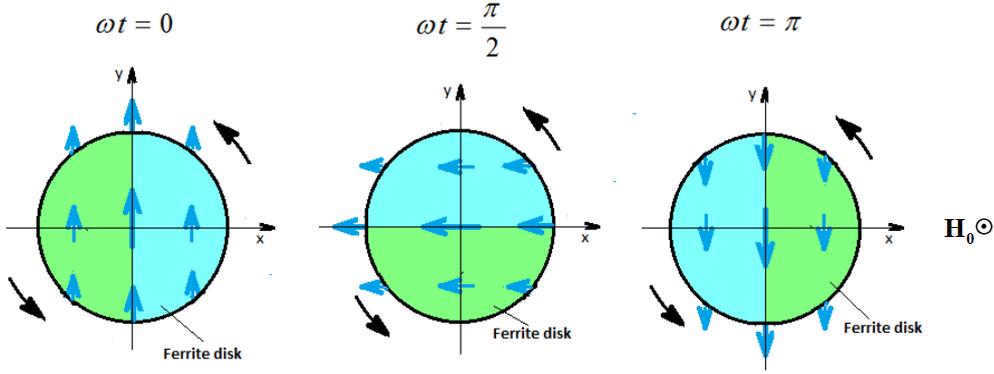

Fig. 4. A schematic representation of the fields of the *RH* and *LH* helices on the disk cross section plane at a bias magnetic field directed along *z* axis. The phases $\omega t$ is a dynamic phase of an external RF EM field. Magnetization dipoles inside and on the lateral border of a disk for the main mode of the MDM spectrum are illustrated with blue arrows. A dipolar configuration of a membrane function $\tilde{\eta}_n(r,\theta)$ is schematically shown as two semicircles of different colors. At the EM-field phase variation of $\omega t = \pi$, one has an additional angle of $\pi$ due to the phase variation of membrane function $\tilde{\eta}_n(r,\theta)$. This means that there is a $2\pi$ phase change in the laboratory frame.

In a rigidly nonrotating ferrite disk, the rotating fields are due to two types of rotating helical MS modes: the *RH* and the *LH* helices. The helices are intertwined without interaction. For the two helices, we have a real half-integer angular momentum, with the 'up' and 'down' states (depending on the direction of the bias field). As a result, for a given direction of a bias magnetic field, the membrane function is presented as a two-component sprinor [see Eq. (9)]. The fundamentally distinctive feature between small-scale exchange-interaction magnons and large-scale MDM magnons is that the former are condensed bosons [17], but the latter are condensed fermions. When the frequency of the orbital rotation of MDM resonances in a ferrite disk is close to the ferromagnetic resonance frequency, the precessing magnetic dipoles become strongly correlated and one observes fermionization of the system composed of bosons. The orbital degrees of freedom would behave as if we have fermions in the system, even though our system is composed of bosons. Due to nonzero circulation of magnetization on a lateral surface of a quasi-2D ferrite disk at the MDM resonances one observes the anapole moment [see Eq. (7)]. A magnetic dipole moving along a border contour $\mathcal{L} = 2\pi\mathcal{R}$ acquires a geometric phase only when the circulation is described by double-valued edge MS wave functions. In this case, the persistent magnon-BEC current is a steady flow of magnetic dipoles that produces an electric field. This magnon current can be viewed as a ring flow arising from the Aharonov-Casher phase [56, 57].

Along with the above condition $\vec{\nabla}\cdot\vec{m}\neq 0$ defining the magnetic current density, the condition $\vec{\nabla}\times\vec{m}\neq 0$ presumes the presence of the electric-current density in macroscopic Maxwell's equations [50, 58, 59]. In a quasi-2D normally magnetized ferrite disk, the magnetization vector $\vec{m}$ has only in-plane components. Due to variations of the magnetization along the *z* coordinate in a ferrite disk, one can detect also in-plane components of the vector $\vec{\nabla}\times\vec{m}$. Using the representation (8), we obtain:

$$\left(\vec{\nabla}\times\vec{m}_n\right)_{r,\theta} = -\frac{\partial(m_n)_\theta}{\partial z}\vec{e}_r + \frac{\partial(m_n)_r}{\partial z}\vec{e}_\theta = -C_n \frac{\partial \xi_n(z)}{\partial z}\left[\left(i\chi_a \frac{\partial \tilde{\varphi}_n}{\partial r} - \frac{\chi}{r}\frac{\partial \tilde{\varphi}_n}{\partial \theta}\right)\vec{e}_r + \left(\chi \frac{\partial \tilde{\varphi}_n}{\partial r} + i\frac{\chi_a}{r}\frac{\partial \tilde{\varphi}_n}{\partial \theta}\right)\vec{e}_\theta\right]$$
(25)

For the main thickness mode [42, 55], the vector $\left(\vec{\nabla}\times\vec{m}_n\right)_{r,\theta}$ is antisymmetric with respect to *z* axis.

The magnetization vector $\vec{m}$ has curl component only when nonzero orbital circulation of magnetization on a lateral surface of a quasi-2D ferrite disk occurs at the MDM resonances. This circulation of vector $\vec{m}$ around a contour $\mathcal{L} = 2\pi\mathcal{R}$ is expressed by double-value MS functions. The presence of both potential and curl component of vector $\vec{m}$ is the cause of appearance of pseudo scalar quantity [42]:

$$\vec{m}\cdot\left(\vec{\nabla}\times\vec{m}\right)^* = -i2C^2 \xi(z)\frac{\partial \xi(z)}{\partial z}\left(\chi \frac{\partial \tilde{\varphi}}{\partial r} + \frac{\chi_a}{r}\nu\tilde{\varphi}\right)\left(\frac{\chi}{r}\nu\tilde{\varphi} + \chi_a \frac{\partial \tilde{\varphi}}{\partial r}\right).$$
(26)

In a quasi-2D ferrite disk, the dipole-dipole interaction couples the precessing electron spin and orbital angular momenta of magnetization. Since, in contrast to the Einstein-de Haas effect [23], we observe the angular momentum transferred not to rotation of a rigid body, the question arises about the stability of the MDMs. This contributes to the spontaneous formation of topological spin textures. In this texture, local torques of magnetization must counteract the azimuth anisotropy that would otherwise destroy the MDM. There should be inherent torques with topological magnetic charges. It can be assumed that in a case of a thin ferrite disk, the right- and left-handed internal helical MS waves, shown in Fig. 1, end with surface magnetic topological charges on the top and bottom planes of the ferrite disk. The *RH* and *LH* helices emerge on the plane surfaces of the disk with the opposite chirality of topological charges. We conventionally refer to them as magnetic charges $\rho_m^{(RH)}$ and $\rho_m^{(LH)}$. These topological charges with spiraling texture were observed numerically in Ref. [42]. This situation is shown schematically in Fig. 5. Magnetization dipole in the center of a disk is shown as a blue arrow. This topological texture of magnetization on the plane surfaces is definitely related to the magnon current on the lateral surface of a ferrite disk. It should be noted, however, that within our analytical model with separation of variables, where we neglect the fields in the corner regions of the ferrite disk [34, 41], a theoretical analysis of the relationships between the magnon currents on the lateral surface and the plane surfaces of a ferrite disk is hardly possible. Further numerical studies of this problem are required.

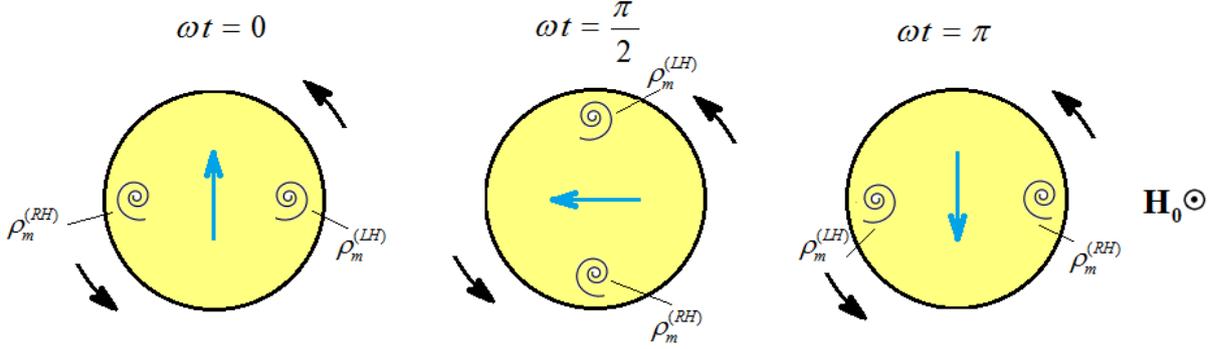

Fig. 5. Topological magnetic charges on a plane surface of a ferrite disk. Magnetization dipole in the center of a disk is shown as a blue arrow. For a two-helix resonance, the topological magnetic charges are on diametrically opposite regions on the disk plane.

## IV. ELECTRIC FIELD STRUCTURES AT THE MDM RESONANCES

Our analysis shows that the spectral problem solution for scalar wave function $\psi(\vec{r},t)$ with the Dirichlet-Neumann boundary conditions are the solutions of the Schrodinger-like equation with energy eigen states. For the EM boundary conditions, one has the solutions in a complex vector space, which represent rotation on vectors in the quantum Hilbert space.

Considering the scalar wave function $\psi(\vec{r},t)$ as a generating function that determines the field dynamics of MDM resonances, we neglect the electric displacement current. Certainly, using the macroscopic Maxwell equation for a magnetic insulator

$$\vec{\nabla} \times \vec{B} = \mu_0 \left( \varepsilon_0 \frac{\partial \vec{E}}{\partial t} + \frac{\partial \vec{p}}{\partial t} + \vec{\nabla} \times \vec{m} \right) = \mu_0 \left( \frac{\partial \vec{D}}{\partial t} + \vec{\nabla} \times \vec{m} \right), \quad (27)$$

we can see that for a magnetic material with strong temporal dispersion at ferromagnetic-resonance frequencies [where we have $\vec{B} = \mu_0(\vec{H}+\vec{m}) \approx \mu_0 \vec{m}$], an inequality

$$\left| \frac{\partial \vec{D}}{\partial t} \right| << \left| \vec{\nabla} \times \vec{m} \right| \quad (28)$$

gives $\vec{\nabla} \times \vec{H} = 0$. In this case, we have the magnetostatic description assuming that a dynamical magnetic field is represented as $\vec{H} = -\vec{\nabla}\psi$.

It is asserted [58], that for a subwavelength magnetic sample with strong temporal dispersion of a material, an inequality (28) means that variations of electric energy is negligibly small compared to variations of magnetic energy. For this reason, in many studies of MS oscillations and waves [43, 58, 60, 61], a role of the electric fields, defined based in the Faraday's law

$$\vec{\nabla} \times \vec{E} = -i\omega\mu_0 \left( \vec{H} + \vec{m} \right),  \qquad (29)$$

had been ignored. This also means that at the MS description of magnetic oscillations, any effects of dynamical electric polarization $\vec{p}$ in magnetic insulator, caused by the electric field $\vec{E}$, should be completely ignored.

Nevertheless, the fact of the presence of the dynamical electric polarization $\vec{p}$ does not contradicts to the MS description of MDM oscillations if we define the induced macroscopic polarization in terms of adiabatic flows of surface magnetic currents in a ferrite disk. Such a description of polarization is closely related to a Berry phase and may be regarded as a multi-valued quantity [62 – 64]. It was shown [38] that in the coordinate frame of orbitally driven field patterns, the lines of the electric field $\vec{E}$ as well the lines of the polarization $\vec{p}$ are "frozen" in the lines of magnetization $\vec{m}$. It means that there are no time variations of vectors $\vec{E}$ and $\vec{p}$ with respect to vector $\vec{m}$ (and, certainly, with respect to space derivatives of vector $\vec{m}$). So, all time variations of the vectors $\vec{E}$ and $\vec{p}$ are synchronized with time variations of vector $\vec{m}$ in a laboratory frame. Since at the MDM resonances in YIG, the MS-potential waves $\psi(\vec{r}, t)$ provide magnetization vectors with spin and orbital rotation, the $\vec{E}$ and $\vec{p}$ vectors have spin and orbital angular momentums too. In such a case, the Faraday law (29) is not in contradiction with the MS description. In a frame of an orbitally rotating coordinate system, the vector $\vec{D}$ is considered constant. Thus, inequality (28) is satisfied.

The polarization, defined by an anapole moment (7), is determined by the modulus of the "polarization quantum of polarization". It is regarded as a multi-valued quantity. The electric polarization of a MDM ferrite disk is closely related to a Berry phase of the MS wave functions. Topological magnetic currents provide us with the possibility to have rotational symmetry by a turn over a regular-coordinate angle $2\pi$ at the $\pi$-shift of a dynamic phase of the external RF EM field. This means that the frequency of the orbital rotation must be twice of the EM wave frequency $\omega$. Due to circulation of a chiral magnetic current on a lateral surface of a ferrite disk, electric charges appear on the top and bottom planes of the ferrite disk.

Based on the MS description ($\vec{\nabla} \times \vec{H} = 0$) and taking into account that YIG is an isotropic dielectric ($\vec{\nabla} \cdot \vec{E} = 0$), we rewrite Eq. (29) as

$$\nabla^2 \vec{E} = i\omega\mu_0 \vec{\nabla} \times \vec{m}.  \qquad (30)$$

As we pointed out above, a magnetization vector $\vec{m}$ has curl component only due to nonzero orbital circulation of magnetization on a lateral surface of a quasi-2D ferrite disk at the MDM resonances. This shows that the electric fields in MDMs are observed exclusively due to the effect of the orbital field rotation. Since is the vector $\left( \vec{\nabla} \times \vec{m}_n \right)_{r,\theta}$ is antisymmetric with respect to $z$ axis, the electric field is antisymmetric with respect to $z$ axis as well.

The electric field can be defined by integration of Eq. (30). For the electric field outside a ferrite disk, we have:

$$\vec{E}_{outside}(\vec{r}) = -\frac{i\omega\mu_0}{4\pi}\left(\int_V \frac{\vec{\nabla}'\times\vec{m}(r')}{|\vec{r}-\vec{r}'|}dV' + \int_S \frac{\vec{m}(r')\times\vec{n}'}{|\vec{r}-\vec{r}'|}dS'\right). \quad (31)$$

Surface integral in Eq. (31) is caused by the orbital circulation of magnetization on the lateral surface of a quasi-2D ferrite disk at the MDM resonances. When analyzing above Eq. (23) for the magnetic field we argued that this is geometric-phase circulation leading to the observation of an anapole moment. The electric field distribution at MDM resonances based on the volume integral in Eq. (31) was analyzed in [37]. Fig. 6 schematically shows such distributions on the top and bottom plane surfaces of a ferrite disk for a given direction of a bias magnetic field. For the MDM quantized states, the electric field has both orbital and spin angular momentums.

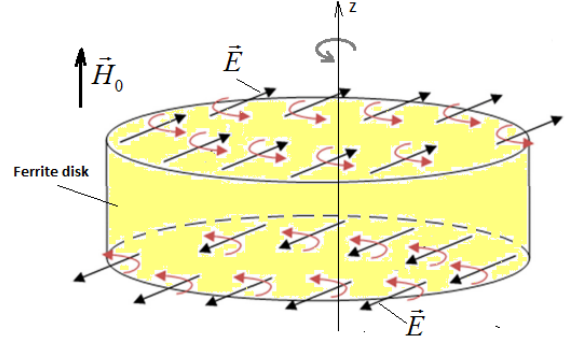

Fig. 6. The MDM electric field distribution on the top and bottom plane surfaces of a ferrite disk for a given direction of a bias magnetic field. The electric field has both orbital and spin angular momentums. The electric field distribution is antisymmetric about the *z* axis. In a crystal of an isotropic magnetic dielectric, the vectors of electric polarization and electric field are mutually parallel.

It should be noted, once again, that the sources of the electric field (31) are not electric charges, but magnetic currents. At the MDM resonances, the electric field and electric polarization arise from the magnetization dynamics in a YIG disk. In the orbitally rotating field patterns, originated from the magnetization dynamics in a ferrite disk, no displacement electric currents are taken into consideration. At the MS description, the effects of quasistatic (time-dependent) electric polarization by the RF electric field should be ignored. In the reference frame co-rotating with the magnetization in a ferrite disk, the electric polarization in is not time varying in any MDM quantum state. In a crystal of an isotropic magnetic dielectric, the vectors of electric polarization and electric field are mutually parallel.

The near field of the disk at MDM resonance is considered as a structure of synchronously rotating magnetic and electric fields. This field structure is called the ME field [37, 42]. The region of the ME field is a subwavelength region in which the EM symmetry is broken. In a laboratory frame, one views a system of "glued" magnetic dipole and electric quadrupole rotating at a microwave frequency. These magnetic dipole and electric quadrupole are shown in Fig. 7 together with the directions of the vectors of the DC magnetization and the anapole moments.

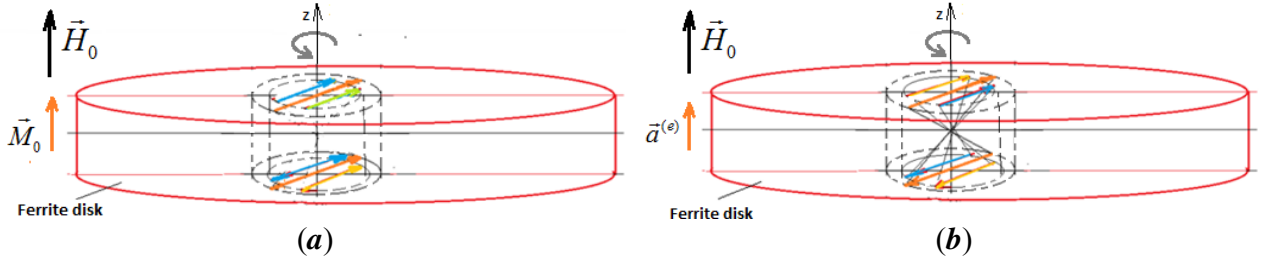

Fig. 7. The structure of magnetic dipole (**a**) and electric quadrupole (**b**) distributions rotating synchronically at the MDM resonances observed in the planes of a ferrite disk at a given direction of a bias magnetic field. The figure also shows the directions of the DC magnetization and the anapole moment. The orbitally rotating fields of the magnetization and electric polarization are due to two types of rotating helical MS modes: the *RH* and the *LH* helices.

## V. MAGNETOELECTRIC NEAR FIELDS

The electric and magnetic components of the ME fields are calculated based on Eqs. (23) and (31). In our description, the magnetic fields of MDMs are potential fields, both inside and outside a ferrite disk. The MDM electric fields are curl fields inside the disk. Outside the ferrite disk the electric fields described by Eq. (31) are potential fields: $\vec{E}_p \equiv -\vec{\nabla}\phi$. At the same time, in the vacuum region outside the ferrite disk, we have also a curl electric field defined as

$$\vec{\nabla} \times \vec{E}_c = i\omega\mu_0 \vec{\nabla}\psi . \tag{32}$$

At the MDM resonance, every mode is characterized by power-flow vortices. The orbital angular-momentum of the power flow density in a near-field region is expressed as

$$\vec{\mathcal{L}}_z = \frac{1}{2}\text{Re}\left[\vec{r} \times \left(\vec{E} \times \vec{H}^*\right)\right] . \tag{33}$$

Depending on a direction of a bias magnetic field, we can distinguish the clockwise and counterclockwise topological-phase rotation of the fields outside the ferrite disk. The direction of an orbital angular-momentum $\vec{\mathcal{L}}_z$ is correlated with the direction of a bias magnetic field $\vec{H}_0$ (along +z axis or –z axis) [37, 42, 54, 65]. Quantized power-flow vortices are observed due to curl electric fields and potential magnetic fields. Since there are no curl magnetic fields, we certainly have no effects of EM retardation in the near-field region.

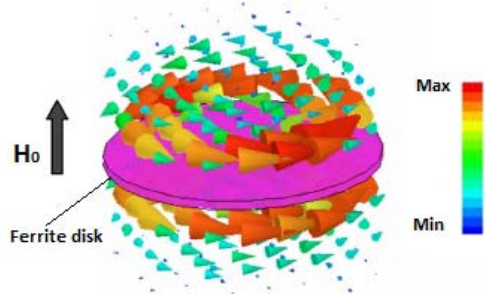

Fig. 8. Power-flow vortices above and below a MDM ferrite disk

It was shown [37] that together with power-flow vortices, in the near-field region adjacent to the MDM ferrite disk, there exists also another quadratic-form parameter determined by a scalar product between the electric and magnetic field components:

$$F = \frac{\varepsilon_0}{4} \operatorname{Im}\left[\vec{E} \cdot \left(\nabla \times \vec{E}\right)^*\right] = \frac{\omega \varepsilon_0 \mu_0}{4} \operatorname{Re}\left(\vec{E} \cdot \vec{H}^*\right). \qquad (34)$$

The vector $\nabla \times \vec{E}$ measures the rotation of the vector field $\vec{E}$. So, a scalar product of $\nabla \times \vec{E}$ with $\vec{E}$ is an indicator of how much the electric-field vector field rotates around itself. We name parameter $F$ as the ME-field helicity density. In a case of ME helicity, this is a time-odd pseudoscalar. This is different from the known concepts of the magnetic, electric and electromagnetic helicities, which are considered as time-even Lorentz pseudoscalars [66, 67]. Pseudoscalar is a quantity that changes its sign when one changes *RH* coordinate system to *LH* coordinate system and vice versa. In the case of EM fields, helicity can be considered as the difference of the number of the *RH* and *LH* propagating photons. For ME fields, the parameter of helicity indicates characteristics of the near-field structure.

At each MDM resonance, we observe time average near-field parameters: the power-flow vortex and the field helicity. An analysis shows that the ME near fields are characterized by unique symmetry properties, both with respect to time reversal and parity. The electric and magnetic fields outside a ferrite disk are rotating fields, which are not mutually perpendicular in vacuum. Depending on a direction of a bias magnetic field, we can distinguish the clockwise and counterclockwise topological-phase rotation of the fields. The mirror symmetry of the field vectors with respect to the plane of symmetry of the disk is very specific. As we can see from the analytical and numerical results [37, 42, 54, 65], the magnetic field vectors appear to be polar relative to the plane of the disk. At the same time, electric field vectors behave like axial vectors with improper rotation. This means that in the case of the electric field vectors, we have a mirror inversion combined with the rotation of the mirror plane around the disk axis. These symmetry properties of the field vectors are illustrated in Fig. 9.

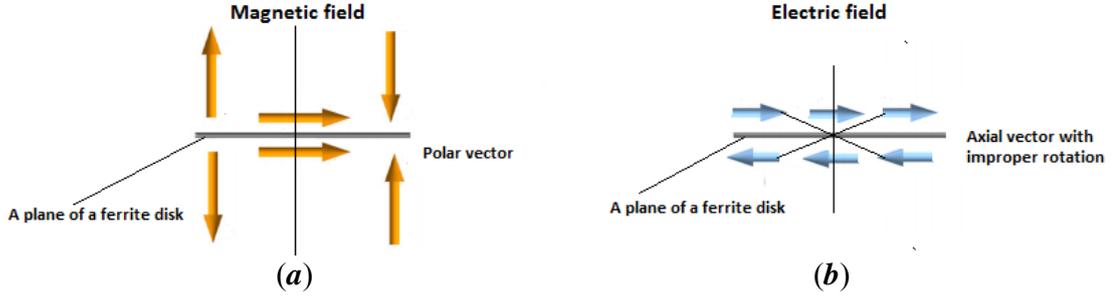

Fig. 9. The symmetry properties of the ME field vectors outside a ferrite disk at the MDM resonance. (*a*) Magnetic field, (*b*) electric field. Both the electric and magnetic fields are divergence free. There are no electric and magnetic charges.

The effect of ME helicity is due to the presence of both the curl and potential electric fields of the modes outside the disk. Parameter *F* appears only at the MDM resonances. A sign of the helicity parameter depends on a direction of a bias magnetic field. Because of time-reversal symmetry breaking, all the regions with positive helicity become the regions with negative helicity, when one changes a direction of a bias magnetic field:

$$F^{\vec{H}_0 \uparrow} = -F^{\vec{H}_0 \downarrow}.\tag{35}$$

The helicity-density distribution is related to the angle between the spinning electric and magnetic fields. Fig. 10 shows the helicity density distributions above and below a ferrite disk at the MDM resonance at two opposite directions of a bias magnetic field. Also, there are shown the electric and magnetic components of the ME field.

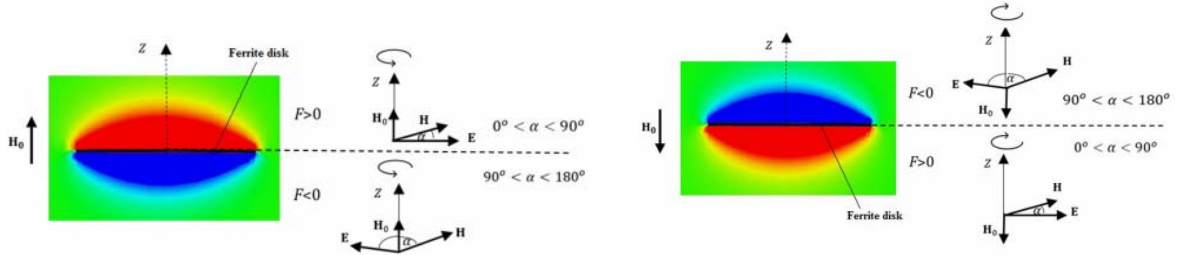

Fig. 10. The helicity density distributions above and below a ferrite disk at the MDM resonance at two opposite directions of a bias magnetic field. The electric and magnetic fields outside a ferrite disk are rotating fields, which are not mutually perpendicular. In a green region *F* = 0: the angle between the electric and magnetic fields is $90°$.

The helicity parameter *F* is a pseudoscalar: to come back to the initial stage, one has to combine a reflection in a ferrite-disk plane and an opposite (time-reversal) rotation about an axis perpendicular to that plane. In a space symmetrical structure, an integral of the ME-field helicity over an entire near-field vacuum region should be equal to zero. Such "helicity neutrality" can be considered as a specific conservation law of helicity.

Since no vorticity is observed in the area of the disk plane adjacent to the disk axis, the potential flow approximation can be applicable. In this region, both potential magnetic and potential electric fields satisfy the Laplace equation. One can rewrite Eq. (34) in following form:

$$F = \frac{\omega \varepsilon_0 \mu_0}{4} \mathrm{Re}\left(\vec{\nabla}\phi \cdot \vec{\nabla}\psi^*\right). \tag{36}$$

At the same time, following the results in Ref. [68], we have:

$$F = \frac{1}{2}\mathrm{Re}\left|\vec{E}\cdot\vec{H}^*\right| = \frac{1}{2}\mathrm{Im}\left|\left[\vec{E}\times\vec{H}^*\right]_z\right|. \tag{37}$$

As a result, one obtains:

$$\mathrm{Re}\left|\left(\vec{\nabla}\phi\cdot\vec{\nabla}\psi^*\right)\right| = \mathrm{Im}\left|\left[\vec{\nabla}\phi\times\vec{\nabla}\psi^*\right]_z\right|. \tag{38}$$

It can be assumed that in a 2D approximation of the ME near field, both the electric and magnetic fields lie in the plane of the disk. In this case, for mode $n$ we have Laplace's equations for membrane functions:

$$\left(\nabla_\perp^2 \psi\right)_n = \frac{\partial^2 \tilde{\varphi}_n}{\partial x^2} + \frac{\partial^2 \tilde{\varphi}_n}{\partial y^2} = 0 \quad \text{and} \quad \left(\nabla_\perp^2 \phi\right)_n = \frac{\partial^2 \tilde{\vartheta}_n}{\partial x^2} + \frac{\partial^2 \tilde{\vartheta}_n}{\partial y^2} = 0. \tag{39}$$

Here: $\tilde{\varphi}_n$ and $\tilde{\vartheta}_n$ are membrane functions of the scalar wave functions $\psi_n$ and $\phi_n$, respectively. Suppose the plane of the ferrite disk is the complex plane: $x + iy$. On this plane, we have a complex number $a + ib$, where $a$ and $b$ are real positive numbers: $\left|\left(\vec{\nabla}_\perp\phi\cdot\vec{\nabla}_\perp\psi^*\right)\right| \equiv a$ and $\left|\left[\vec{\nabla}_\perp\phi\times\vec{\nabla}_\perp\psi^*\right]_z\right| \equiv b$. From Eq. (39) it follows that $a = b$. It means that on the central region of the disk complex plane, we have certain circles of 2D potential flows. We can say also that in the central region of the disk, the helicity $F$ is the result of a dot product between a polar vector (the magnetic field) and pseudovector (the electric field). Conservation of the ME potential energy means that we have helicity neutrality: $F(+z) = -F(-z)$. This allows us to write that for any mode $n$:

$$\mathrm{Re}\left[\left(\vec{\nabla}_\perp\phi\right)_n \cdot \left(\vec{\nabla}_\perp\psi^*\right)_n + \left(\vec{\nabla}_\perp\phi^*\right)_n \cdot \left(\vec{\nabla}_\perp\psi\right)_n\right] = 0. \tag{40}$$

Suppose now that we have a topologically trivial case: $F = 0$. It means that

$$\mathrm{Re}\left(\vec{E}\cdot\vec{H}^*\right) = \mathrm{Re}\left(\vec{\nabla}_\perp\phi\cdot\vec{\nabla}_\perp\psi^*\right) = 0. \tag{41}$$

According to the Cauchy–Riemann equations, the fact that the dot product of two gradients in Eq. (41) is zero implies that the gradient of $\tilde{\vartheta}_n$ (that is, the in-plane electric field) must point along the curves $\tilde{\varphi}_n$ and thus perpendicular to the in-plane magnetic field.

## VI. CAVITY QUANTUM ELECTRODYNAMICS WITH MDM DISKS. CASIMIR TORQUE

Compared to other well-known mechanisms of interaction of photons with dipole-carrying excitations, in this paper we propose a mechanism of such an interaction via the torque transfer. The presence of rotating resonances of MDM eigenmodes in a single-mode cavity means that specific field structures with discrete states of angular momentum must exist between the ferrite disk and the cavity walls, separated by a certain distance in vacuum. This assumes the conservation of angular momentum (see Fig. 11). In order to have a balance of an angular momentum, the cavity body will tend to rotate in the opposite direction to the direction of rotation of the MDM oscillations. When the rigid-body rotations of the disk and cavity are prohibited, the wavefronts in vacuum will be curved due to the action of the Casimir torque.

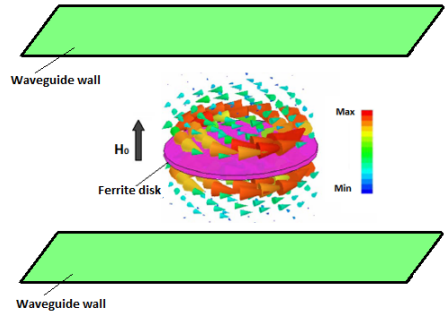

Fig. 11. In the ferrite disk - cavity system, the conservation of angular momentum should be observed at MDM resonances.

If a ferrite particle with MDM resonances is under interaction with a microwave cavity, the states of this cavity object change. The character and value of these changes depend on the MDM quantized states and so can serve as its qualitative characteristics. The microwave measurement reflects interactions when the ferrite disk is with respect to two external parameters – a bias magnetic field $H_0$ and a signal frequency $\omega$. In neglect of losses, there should exist a certain uncertainty limit stating that [38]

$$\Delta f \Delta H_0 \geq \text{uncertainty limit}. \qquad (42)$$

This uncertainty limit is a constant value that depends on the parameters of the disk geometry and the properties of the ferrite material (for example, saturation magnetization).

Mathematically, the uncertainty relation between the frequency and the bias magnetic field arises from the fact that the MS scalar wavefunction of MDMs $\psi$ can be expressed in the two corresponding orthonormal functional bases. These orthonormal functional bases are shown in Fig. 12.

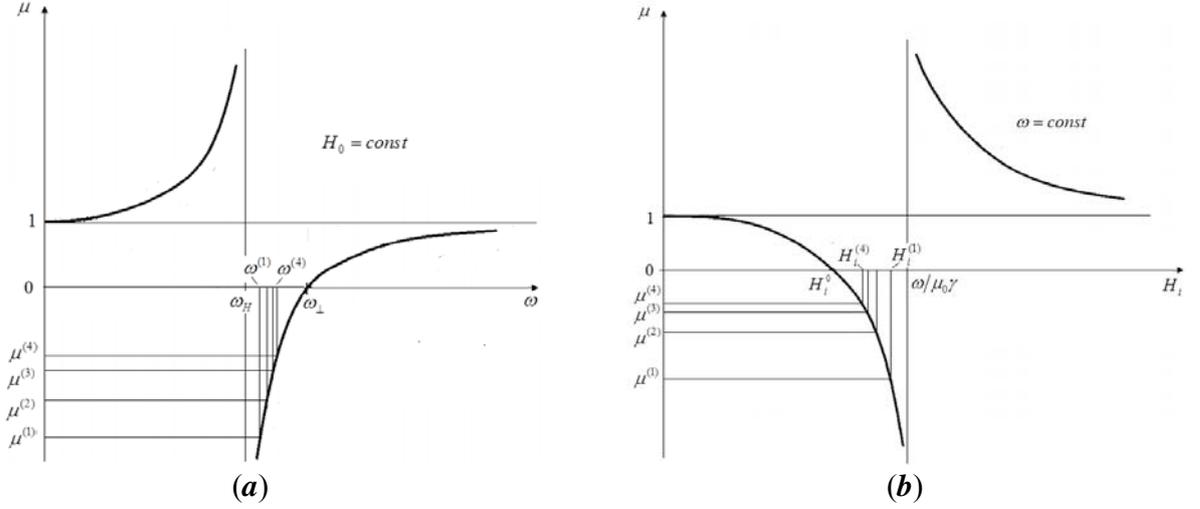

Fig. 12. Two corresponding orthonormal functional bases shown for the first four MDM resonances. (*a*) Diagonal component of the permeability tensor $\mu$ versus frequency at a constant bias magnetic field, (*b*) versus a DC internal magnetic field at a constant frequency. The discrete quantities of $\mu$ are negative. The frequency range is $\omega_H < \omega < \omega_\perp$. The bias field region is $H_i^\diamond < H_i < \omega/\mu_0\gamma$, where $H_i^\diamond \equiv \sqrt{\left(\dfrac{\omega}{\mu_0\gamma}\right)^2 + \left(\dfrac{M_0}{2}\right)^2} - \dfrac{M_0}{2}$, $\gamma$ is the gyromagnetic ratio and $M_0$ saturation magnetization.

On can assume that $\psi(\omega)$ is the Fourier transform of $\psi(H_0)$. In such a case, $\omega$ and $H_0$ are considered as conjugate variables. This means that the precision of the position $\omega$ is improved, by using a series of MDMs, thereby weakening the precision of the $H_0$. It implies also that no MDM eigenstate can simultaneously be both an eigenstate of frequency and an eigenstate of a bias magnetic field. It should be noted, however, that since analytical solutions for MDM wavefunction were obtained with definite assumptions [33, 34], it is difficult to prove theoretically that $\omega$ and $H_0$ are canonically conjugate variables.

In Fig. 13, we show the relationship between quantized states of microwave energy in a cavity and magnetic energy in a ferrite disk. At the cavity resonance frequency $f_{cavity\ res}$, a multiresonance spectrum of modulus of the reflection coefficient is related to microwave energy accumulated in the cavity. At MDM resonances, there are jumps of electromagnetic energy, $w_{RF}^{(n)}$. The multiresonance spectrum is observed at a variation of a bias magnetic field, provided that the frequency of each MDM resonance is equal to the cavity resonance frequency:

$$f_{MDM\ res}^{(n)} \equiv f_{cavity\ res}. \tag{43}$$

The condition (43) is satisfied when the MDM resonance peak is located on the top of the cavity-resonance curve. Choosing, as an example, the peak of the first MDM resonance, we show in Fig. 13 that on the top of the cavity-resonance curve, the Fano line shape of a MDM resonance collapses at $H_0^{(1)}$. The scattering cross section of a single Lorentzian peak at the state $f_{MDM\ res}^{(1)} = f_{cavity\ res}$ and $H_0 = H_0^{(1)}$ corresponds to a pure dark mode. This is the lower energy state of the internal energy of the ferrite disk without the orbital rotation of MDM oscillation. In this energy eigenstate, there is no power-flow vortex and the field structure does not depend on the direction of the bias magnetic field.

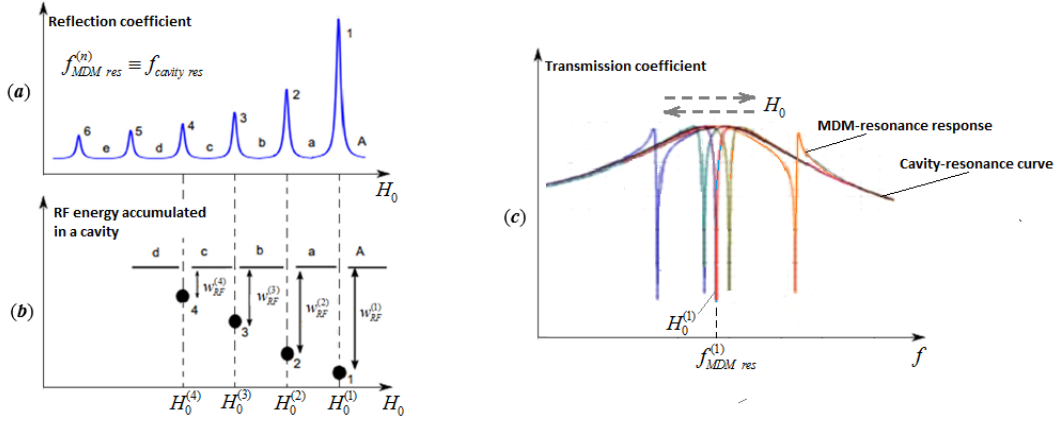

Fig. 13. The relationship between quantized states of microwave energy in a cavity and magnetic energy in a ferrite disk. (*a*) A typical multiresonance spectrum of modulus of the reflection coefficient. (*b*) Microwave energy accumulated in a cavity; $w_{RF}^{(n)}$ are jumps of electromagnetic energy at MDM resonances. (*c*) By the example of the peak of the first MDM resonance, the change in the shape of the Fano resonance with a change in the bias magnetic field is shown. The "blue" and "red" shapes of the Fano-resonance curves are mutually asymmetric. When the direction of the bias magnetic field changes, the "blue" and "red" shapes of the Fano-resonance curves swap places with respect to the cavity resonance frequency. On the top of the cavity-resonance curve, the Fano line shape of the MDM resonance is completely damped at $H_0^{(1)}$. The scattering cross section of a single Lorentzian peak corresponds to a pure dark mode.

When, at the MDM resonances in a cavity, quantization of the cavity vacuum fields occurs by a DC magnetic field at a constant frequency, the internal energy of a ferrite disk decays to lower energy states by the emission of microwave radiation. This would cause the structure to be unstable and implies the existence of negative energy states. The transition to lower energy states is due to the Casimir torque. In structures with rotational symmetry breaking, quantum fluctuations of electromagnetic waves and boundary conditions imposed by the interacting samples causes the samples to rotate to a position of minimum energy. Vacuum fluctuations are expected to generate a Casimir torque [69 – 73]. In our case, the energy of rotational oscillation is due to orbital angular momenta of the power-flow vortices. We argue that interactions between a MDM particle and a waveguide wall is described in terms of the exchange of virtual photons. These vacuum fluctuations, caused by the uncertainty relation (42), we call ME photons. Energy and angular momentum of virtual ME photons are quantized. To satisfy the uncertainty principle, energy of oscillation must be greater than the minimum of the magnetic potential well.

In experiments [28, 29], it was shown that MDM oscillations in a microwave cavity are well excited when a thin ferrite disk is placed in a region of maximum of the RF magnetic field oriented in a disk plane. Obviously, this is the region where the cavity RF electric field, normal to the disk plane, passes through zero. Since in this place the cavity RF electric field is perpendicular to vectors of MDM eigen electric dipoles, we observe quantized states of the "electric" Casimir torque. The time-average "electric" Casimir torque acting on the MDM particle is defined as

$$T_{El}^{(n)} = \left\langle \vec{E}_{extern} \times \vec{P}^{(n)} \right\rangle \cdot \vec{e}_z. \qquad (44)$$

Casimir torques acting on the MDM particle in a cavity are shown in Fig. 14. Obviously, at the MDM eigenstate, the resulting Casimir torque in a rigidly nonrotating disk is zero.

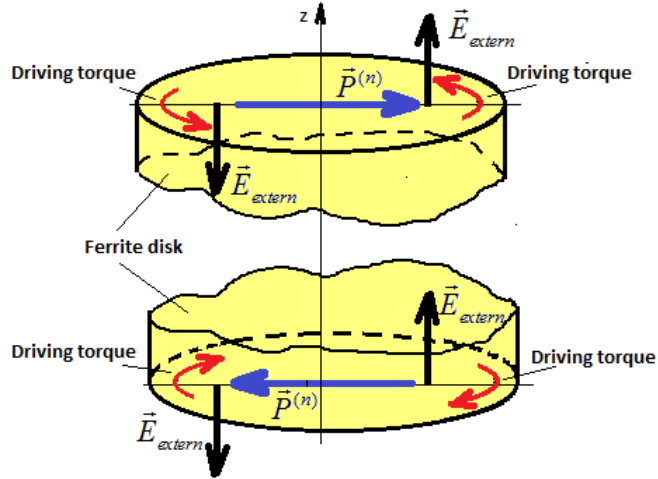

Fig. 14. Casimir torque acting on the MDM particle in a cavity. The vectors of the electric field and polarization are shown on the plane surfaces of the upper and lower halves of the ferrite disk. The disk is placed in the region where the cavity RF electric field, normal to the disk plane, passes through zero. Black arrows are directions of a RF external electric field. Electric polarization of a MDM mode is shown by a blue arrow.

To analyze the topological properties of rotating fields in a rigidly nonrotating disk, consider a TE-mode rectangular-waveguide cavity with a ferrite sample placed in a region of a maximal RF magnetic field. At the location of the disk, the RF electric field perpendicular to the plane of the disk must pass through zero. The disk is considered as a subwavelength particle. At excitation of the MDMs with the orbital rotation of the fields and power-flow vortices, we have also the field rotation in a vacuum region. On a metal wall, rotating electric fields are generated by surface electric currents with spin and orbital moments. Also, topological electric charges should appear on the wall. The region of these surface electric currents and charges on a metal wall is the region of subwavelength singularity. It is obvious that to create this singularity at the MDM resonance, an entire field structure of the microwave cavity should be involved. This is the effect of a strong-coupling regime of MDM polaritons. In the cavity regions where the external RF electric fields, normal to the disk plane, pass through zero, we observe noncontact transfer of angular momentum. This is a quantized effect of angular momentum pumping and emission of twisted photons. It is important to note that in the laboratory frame we observe the situation when the $\pi$ phase change of the cavity field on the waveguide wall is correlated with the $2\pi$ phase change of the orbitally rotating fields in ferrite disk. This should result in the field curvature in vacuum. The presence of curvilinear wavefronts is an energetically favorable regime at the MDM resonance. The effect is related to the local topological charges both on the disk and on the walls [42]. The interaction energy depends on the angular orientation of the MDM eigen electric dipoles and the tangential electric fields inducted on a metal wall, resulting in a torque that would cause their axes to align properly. Assuming that there are no electromagnetic retardation effects above and below a ferrite disk, we observe that the tangential electric fields induced on the metal wall are with the same magnitude as the MDM tangential electric fields on the disk surface, but oppositely oriented. This is possible, however, if metal walls of the cavity are with some ohmic losses. In a vacuum cylinders above and below a ferrite disk, we observe curved quasi-electrostatic modes. These modes appear due to topological charges with spiraling texture on the disk plane and on the metal wall. Such topological charges were observed numerically in Ref. [42]. A sketch of the field structure in the cross section of cavity with an embedded MDM ferrite disk is shown in Fig. 14.

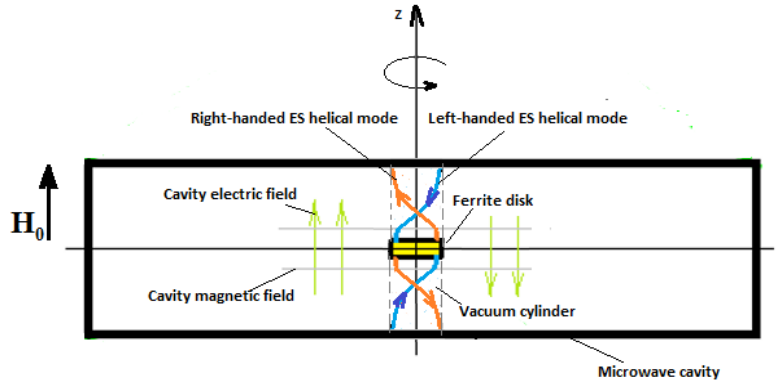

Fig. 14. Sketch of the field structure in the cross section of cavity with an embedded MDM ferrite disk. The disk is put in a region with a maximal RF magnetic field of the TE wavegiude mode. In this area, the RF electric fields, normal to the disk plane, passes through zero. Since the $\pi$ phase change of the cavity field on the waveguide wall should be correlated with the $2\pi$ phase change of the orbitally rotating fields in ferrite disk, the ME photons are viewed as double-helix resonances of quasi-electrostatic modes in vacuum regions above and below the disk. The curved electric fields in vacuum cylinders are highlighted with the blue and red arrows.

In the above analysis, we considered ME virtual photons, assuming that in the subwavelength spaces above and below a ferrite disk there are no EM retardation effects. When the EM retardation processes in these regions cannot be ignored, interaction between ME and EM photons occurs. Will the wavefronts be curved in this case of interacting EM and ME photons? Numerical studies in Refs. [65, 74] provide a very nontrivial answer to this question. The structure analyzed in [74] is shown in Fig. 15 (*a*). The ferrite disk is embedded in a dielectric cylinder, which is placed between two metallic walls of a microwave waveguide. The dielectric constant of the dielectric cylinder is high enough ( $\varepsilon_r = 100$ ) for electromagnetic resonance to occur between the metal walls. However, together with standing electromagnetic waves along the *z* axis, we observe states of the ME field in a dielectric cylinder. These states of the ME field (indicated by the regions with nonzero helicity parameters *F*) have a periodicity correlated with the distribution of EM waves along the *z* axis (see Fig. 15 (*b*)). Fig.

15 (*c*) depicts the structure of the electric field on the surface of the dielectric cylinder. This is the double-helix-resonance structure that rotates orbitally around the *z* axis.

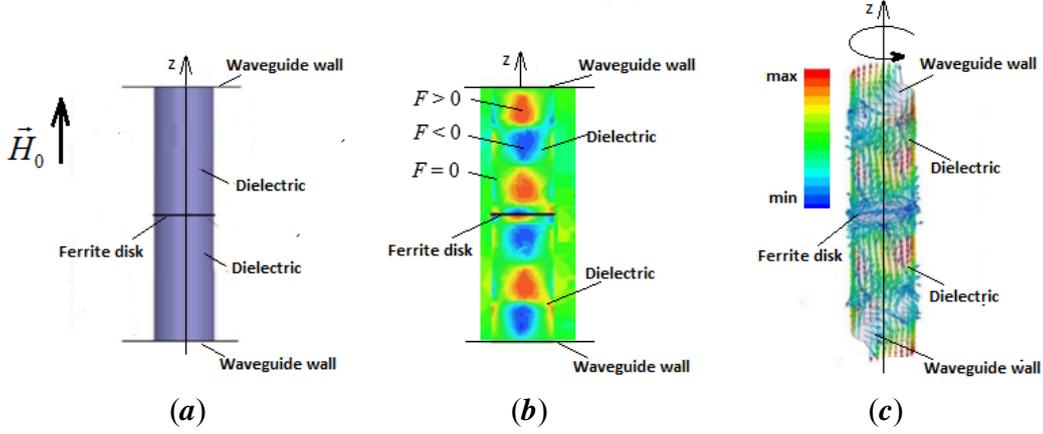

Fig. 15. Interaction between ME and EM photons in a dielectric rod with an embedded MDM ferrite disk. Helical waves. (*a*) The microwave structure. (*b*) Distribution of the helicity parameters in dielectrics. (*c*) Twisted magnon-polariton, depicted by the structure of the electric field on the surface of the dielectric cylinder.

The observed effects of the field curvature are caused by topological phase changes over the scale of the wavelength along the EM-wave path. This leads to bending the EM radiation in unusual ways.

**VII. ROTATIONAL SUPERRADIANT SCATTERING OF MICROWAVE PHOTONS BY MDM VORTICES**

In studies of MDM spectra in a single-mode microwave cavity, we use a constant signal frequency with a variation of a bias magnetic field. At the same time, the MDM spectral properties can be observed in the propagation-wave behavior in a microwave waveguide when bias magnetic field is constant and the signal frequency is changed. In this case, one observes the effect of rotational superradiant scattering of microwave photons by MDM vortices. This effect was verified numerically in of our previous works [54, 65, 74]. Now we are trying to take a fresh look at this phenomenon. For this purpose, it is pertinent to adduce a few pictures from our former papers.

In Fig. 16, we present the electric-field structures of an EM wave propagating in a waveguide with an embedded ferrite disk. The bias magnetic field is directed normally to the disk plane. The fields are shown on a vacuum plane above the ferrite disk. In Fig. 16 (*a*), we see an undistorted wavefront at the frequency outside the MDM resonance. In Fig. 16 (*b*) we show a distorted wavefront at the MDM resonance frequency.

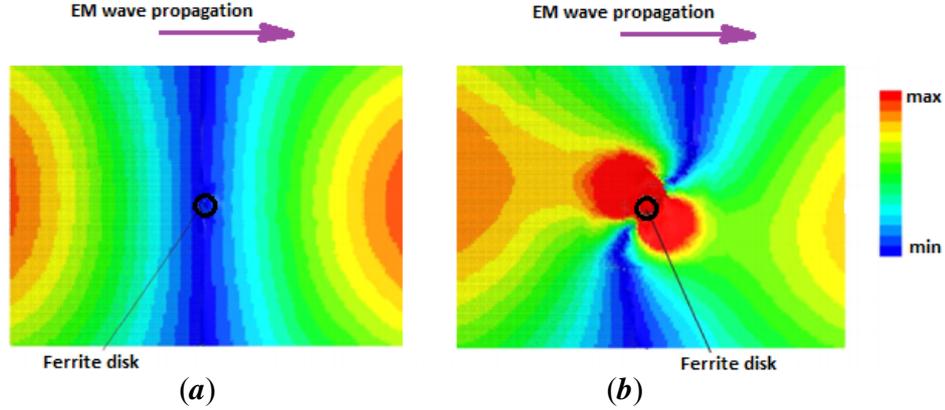

Fig. 16. The electric-field structures of an EM wave propagating in a waveguide with an embedded ferrite disk. (*a*) Undistorted wavefront at the frequency outside the MDM resonance. (*b*) Distorted wavefront at the MDM resonance frequency.

A more detailed picture, presented in Fig. 17, confirms the fact that at the MDM resonance frequency, the normal component of the external RF electric field passes through zero at the center of the ferrite disk in any phase.

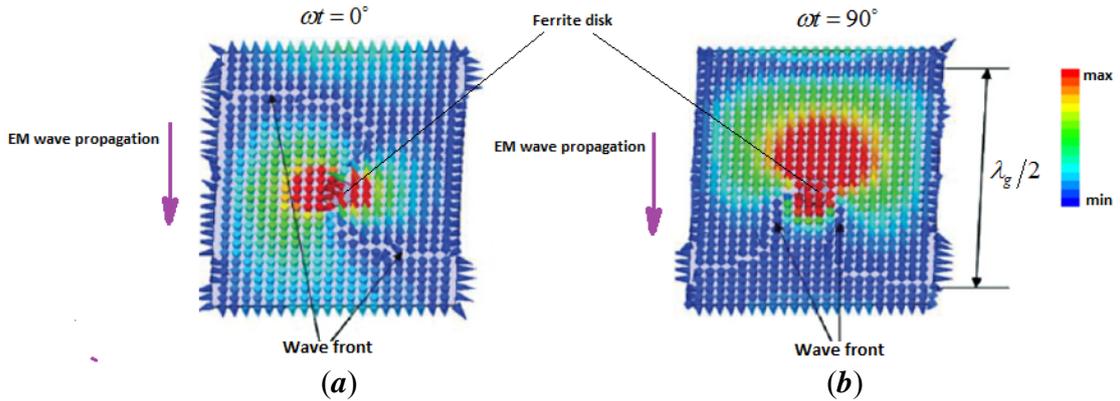

Fig. 17. Inclined top views of the electric field distributions at the MDM resonance frequency on a vacuum plane above the ferrite disk for the two phases. $\lambda_g$ is the wavelength of the TE mode in a rectangular waveguide.

It is important to note here that Fig. 16 (*b*) indicates a relativistic effect that imprints the orbital angular momentum on the EM wave of the waveguide. This is similar to the effects of twisting of light around rotating black holes. In curved spacetime geometries, the direction of a vector is generally not preserved when parallel-transported from one event to another, and light beams are deflected because of gravitational lensing. If the source of the gravitational field also rotates, it drags spacetime with it, causing linearly polarized EM radiation in vacuum to undergo polarization rotation similar to the Faraday rotation of light in a magnetized medium [75].

In a case of ME photons, we observe strong enhancing of the amplitudes of both electric and magnetic fields in the vacuum near-field region [65]. Together with the subwavelength concentration of EM energy, we also view in vacuum subwavelength quadratic forms expressed as $\mathrm{Re}\left(\vec{E}\times\vec{H}^*\right)$

and $\mathrm{Re}\left(\vec{E}\cdot\vec{H}^*\right)$. In the above analysis, we pointed out that to have rotational symmetry, the topological magnetic currents must turn over a regular-coordinate angle $2\pi$ at the $\pi$-shift of a dynamic phase of the external RF EM field. This means that to observe the orbital angular momentum, the frequency of the orbital rotation should be twice the frequency of the EM wave $\omega$. Moreover, we argue that due to double-helix-resonance effects, the wavefronts in vacuum can be curved. Whether these subwavelength effects can be classified also as the effects of nonlinear electrodynamics in vacuum? The effects of nonlinear vacuum electrodynamics are most clearly pronounced in a strong electromagnetic field of astrophysical compact objects such as pulsars. One of the manifestations of vacuum nonlinear electrodynamics effects is curvature of photon trajectories in pulsars [76, 77]. The electromagnetic field equations in nonlinear electrodynamics in vacuum are known to be a system of nonlinear first-order partial differential equations whose general integration methods are still lacking. However, in some cases of a sufficiently simple topography, it is possible to replace the nonlinear action of external fields by introducing tensors of dielectric and magnetic penetrabilities and thus pass to a linear problem of macroscopic electrodynamics. In our studies, ME properties of the near field structure are due rotating electric and magnetic dipoles and can be expressed by introducing tensors of dielectric and magnetic penetrabilities in localized vacuum regions. This leads to the fact that the wavefronts in vacuum can be curved.

An entire structure of an EM-wave waveguide with an embedded MDM ferrite disk can be viewed as shown in Fig. 18. The structure consists of a localized quantum system (the MDM ferrite disk), which is embedded within an environment of scattering states (the microwave waveguide). The coupling between MDM ferrite disk and a microwave waveguide is regulated by means of ME-field vacuum regions.

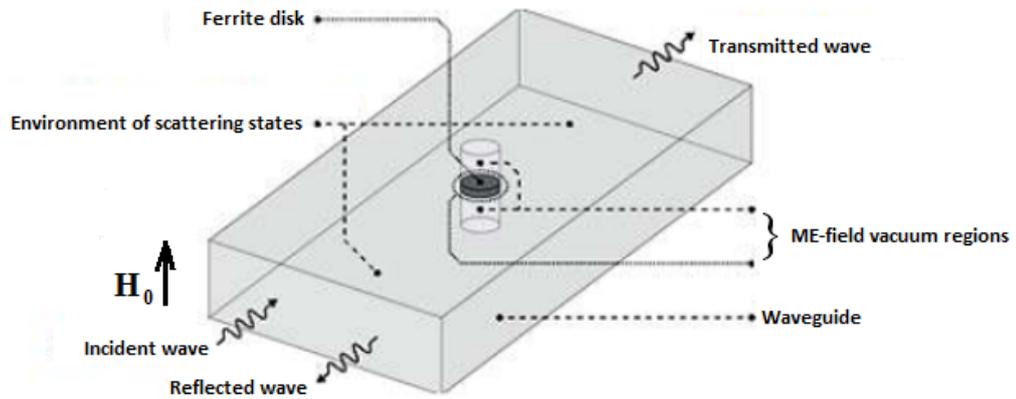

Fig. 18. An interaction of a MDM ferrite disk with a microwave waveguide. The structure consists of a localized quantum system (the MDM ferrite disk), which is embedded within an environment of scattering states (the microwave-waveguide fields). The coupling between MDM ferrite disk and a microwave waveguide is regulated by means of ME-field vacuum regions.

For the structure in Fig. 18, three types of dispersion characteristics can be distinguished. The branches of these characteristics are shown in Fig. 19. A regular EM mode in a waveguide is defined by frequency $\omega$ and a wavenumber $k_{EM}$. At the region $k'_n \gg k_{EM}$ and the frequency around $\omega$, we observe quasi-MS modes of the MDM resonances. The frequency of the orbital rotation should be twice the frequency of the EM wave $\omega$. At the frequency around $2\omega$, there is a certain topological-phase branch of evanescent photon (ME-photon) modes in a waveguide.

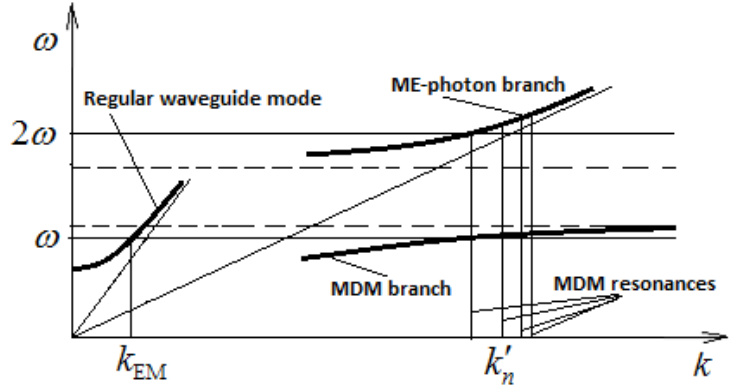

Fig. 19. Due to induced electric currents on a waveguide metal wall, we have the angular-momentum balance. In the presence of these currents, evanescent modes in waveguide appear at frequencies around $2\omega$ in the region $k'_n \gg k_{EM}$. This is a topological-phase branch of ME-photon modes. At the same time, for frequencies around $\omega$, one observes a regular waveguide mode and quasi-MS modes at the MDM resonances in the region $k'_n \gg k_{EM}$. It should be noted that here $k$ is not a linear wavenumber. It rather means spatial scales in the wavenumber space.

The above analysis we illustrate by some numerical results from [65] on the distributions of the electric field in a waveguide with a ferrite disk. In Fig. 20, we show the *xy*-plane components of the electric field scattered by a MDM ferrite disk in a TE-mode waveguide. A bias magnetic field is directed along the *z* axis. The electric fields are shown on a vacuum plane above the disk for two time phases.

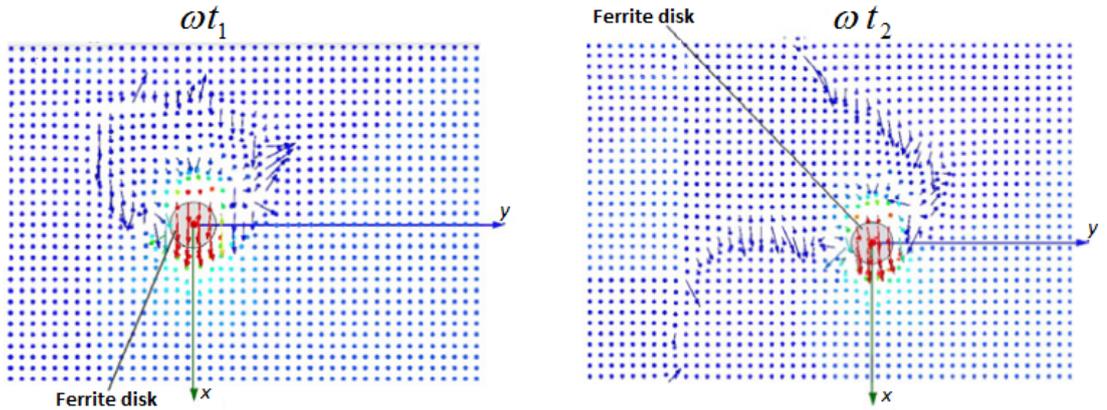

Fig. 20. The *xy*-plane components of electric fields scattered by a ferrite disk at the MDM resonance.

In Fig. 20 clearly visible the contours on which the electric field vectors lie in the *xy* plane. These contours are wave fronts of the waveguide mode. The normal components of the electric-field vectors (shown as blue points in Fig. 20) have different *z*-directions on both sides of the contour.

Schematically, these contours are represented in Fig. 21 for the same time phases as in Fig. 20. In addition, we also showed the wavefront contour for the phase $\omega t_2 + 180°$. Importantly, the spatial orientations of the in-plane electric field vectors on the wavefronts are almost the same as on the disk surface. This means that the phase velocity of "propagation" of topological phases along the wavefronts is extremely small compared to the EM-wave phase velocity. This is imaged by the branch of ME photons shown in Fig. 19. The main distinguishing feature of the wavefront geometry is that each wavefront passes through an MDM ferrite disk. This means that each wavefront is "modulated" (in spatial scale and topology) by MDM resonances. On the other hand, wavefronts trace the geometry of waveguide walls.

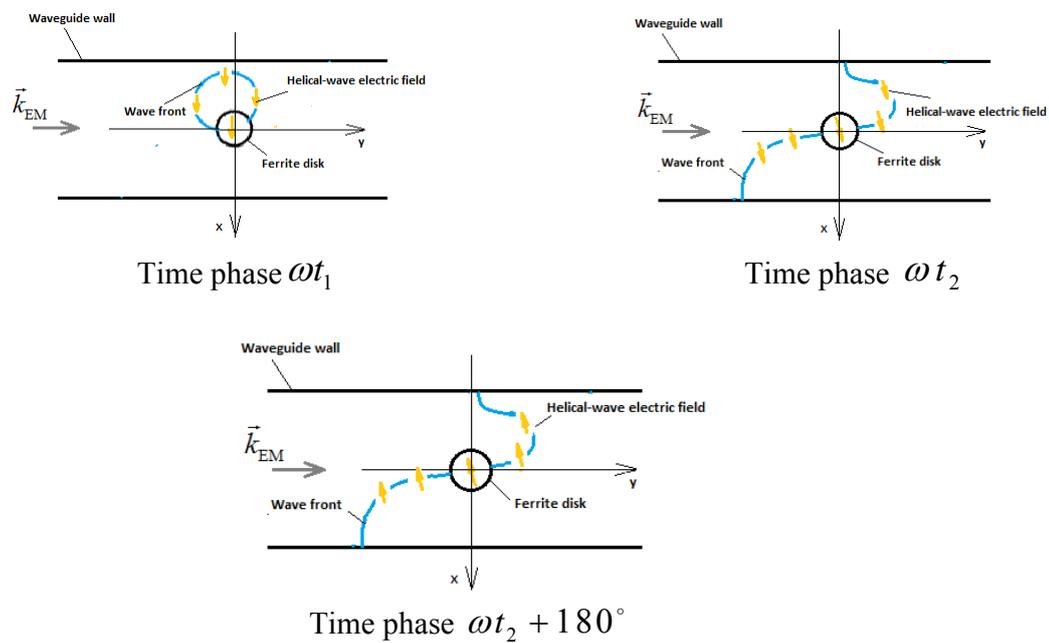

Fig. 21. Schematic representation of wavefronts and topological-phase electric fields at different time phases. Yellow arrows show the spatial orientations of the in-plane electric field vectors.

For a case of a symmetrical position of a ferrite disk related to the waveguide walls, the pictures of *xy* cut, shown on Figs. 20, 21 are asymmetric with respect to the *z* axis. Fig 22 shows wavefronts and topological-phase electric fields on vacuum xy planes situated above and below the disk planes in a region near a ferrite disk.

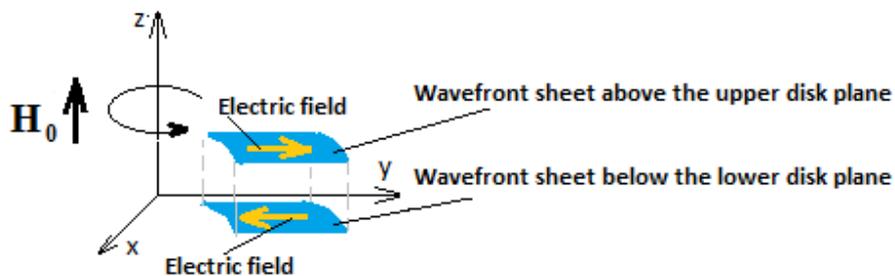

Fig. 22. Sheets of the cut of wavefronts and topological-phase electric fields on vacuum *xy* planes situated above and below the disk planes in a region near a ferrite disk.

The internal motion of magnetization at the MDM resonance is involved in external wave dynamics and interactions. The observation of extremely large wavenumbers of topological photons (ME-photons) in vacuum is possible when the topological-phase rotation occurs at the frequencies higher than the EM wave frequency $\omega$. In the MDM resonance, the orbital rotation of the fields is at the frequency $2\omega$. This is the condition for rotational superradiance. Generation of waves by a rotating body is a well know effect [78]. It was suggested that rotational superradiance leads to the extraction of energy from a black hole [79]. Laboratory detection of this phenomenon in other types of systems is well known [78]. When an incident wave scatters from an obstacle, it is partially reflected and partially transmitted. If the obstacle is rotating, waves can be amplified in the process, extracting energy from the scatterer. In our case, incident EM waves are amplified by extracting energy from the collective motion of orbitally rotating spinning magnetic dipoles. At the MDM resonances, transfer between angular momenta of the magnetic insulator and the microwave structure, demonstrates generation of vortex flows with fixed handedness.

EM waves in vacuum carry the topological phases of MDM resonances. In a 3D representation, the wavefront in vacuum looks like a spiral curve emerging from the MDM ferrite disk and rotating around it. For each time phase $\omega t$, we have different geometrical forms of the wavefront. In the entire structure of an EM-wave waveguide with an embedded MDM ferrite disk, we view that the threads of the electric-field vectors are pulled together to the subwavelength region of the ferrite disk location. These threads are twisted and braided.

**VIII. DISCUSSION AND CONCLUSION**

In our studies, we will refer to new and important aspects of twisted s*ubwavelength* fields. This presumes the existence of specific near fields with unique symmetry properties. Recently, such field structures, called ME fields, were found as the near fields of a quasi-2D subwavelength-size ferrite disk resonator with MDM oscillations. The near fields of these oscillations are characterized by *subwavelength power-flow vortices* with strong enhancing the field intensity.

Generally, it is assumed that topological order is protected from its environment. We study topological MDM resonances coupled to an electromagnetic environment. Our results show that EM waves in vacuum can carry the topological phases of MDM resonances. At the MDM resonances, the superposition of magnon and microwave photons is considered as magnon-polariton condensate. This structure is realized due to magnon condensation caused by magnetic dipole-dipole interaction. The MDM magnon-polaritons have nontrivial topology. Chiral symmetry inherent to the MDM resonances is preserved and protects the topological phase. The magnon-photon wave function is characterized by a nonzero Berry curvature. In the MDM resonances, Berry phase of the MS magnonic wave function has an effect of orbital electric polarization. Such electric polarization cannot be determined from the electric charge and electric current densities in the bulk of an YIG crystal at all. This defines ME properties of the MDM resonances. Due to phase changes of MDM oscillations over the scale of the wavelength along the EM-wave path we observe bending the EM radiation in unusual ways. We have phenomena which is similar to the effects of twisting of light around rotating black holes.

ME fields arising from magnetization dynamics at MDM resonances are pseudoscalar fields. Such fields, which transform as pseudoscalars in the spatial reflection *P* and odd under time reversal *T*, appear as the fields of axion electrodynamics. In axion electrodynamics, the coupling between an axion field and the electromagnetic field is expressed by an additional term in the ordinary Maxwell Lagrangian. This additional term, containing the pseudoscalar axion field and the scalar product $\vec{E} \cdot \vec{B}$, breaks the duality symmetry [81 – 83]. The effect of coupling between the electric field and magnetic fields in a vacuum subwavelength region of the ME-field structure, should be characterized by $\mathcal{L}_{dual+axion}$ Lagrangian.

In a microwave structure with an embedded MDM ferrite disk, the field vectors constitute a composition of threads, which are pulled together to the subwavelength region of the ferrite disk location. These threads are twisted and braided. The spatial orientations of the in-plane electric field vectors on the wavefronts of EM radiation are almost the same as on the disk surface. The phase velocity of "propagation" of topological phases along the wavefronts is extremely small compared to the EM-wave phase velocity. This is manifestation of an effect of nonlocality. It can be assumed that quantized waves propagating along the curved wavefronts of EM radiation at the MDM resonances are the waves of axion fields. Thus, in the entire microwave structure we observe the effect of wave propagation in axion electrodynamics.

Our research is concentrated on manifestations of spin-orbit-type interactions between intrinsic and extrinsic degrees of freedom of electromagnetic waves and quantized MDM oscillations. The goal is to develop a unifying theoretical approach based on fundamental geometric-dynamic effects that describing specific features of behavior of spins and vortices evolving in external fields.